\documentclass[aps,prd, tikz, final,letterpaper, nofootinbib]{revtex4}
\usepackage{textcomp, gensymb}
\usepackage{amsmath, amssymb, amsthm, dsfont, enumerate, eucal, fancyhdr, feynmf, graphicx, mathtools, multirow, physics, tensor, tikz, verbatim}
\usepackage{epstopdf}
\usepackage[many]{tcolorbox}
\usepackage{tikz-3dplot} 
\usetikzlibrary{patterns,perspective}

\usepackage{tikz-feynman}
\tikzfeynmanset{compat=1.0.0}
\tikzset{graviton/.style={decorate, decoration={snake, amplitude=.4mm, segment length=1.5mm, pre length=.5mm, post length=.5mm}, double}}
\usepackage{hyperref}
\begin{document}

\title{Parity-Violating Trispectrum from Chern-Simons Gravity}
\author{Cyril Creque-Sarbinowski}
\email{ccreque@flatironinstitute.org}
\affiliation{Center for Computational Astrophysics, Flatiron Institute, New York, New York 10010, USA}
\author{Stephon Alexander}
\email{stephon_alexander@brown.edu}
\affiliation{Brown Theoretical Physics Center and Department of Physics, Brown University, 182 Hope Street, Providence, Rhode Island, 02903}
\author{Marc Kamionkowski}
\email{kamion1@jhu.edu}
\affiliation{William H. Miller III Department of Physics and Astronomy, Johns Hopkins University, 3400 N. Charles St., Baltimore, Maryland, 21218, USA}
\author{Oliver Philcox}
\email{ohep2@cantab.ac.uk}
\affiliation{Center for Theoretical Physics, Department of Physics,
Columbia University, New York, NY 10027, USA}
\affiliation{Simons Society of Fellows, Simons Foundation, New York, NY 10010, USA}
\date{\today} 

\begin{abstract}
\noindent	
A potential source for parity violation in the Universe is inflation. The simplest inflationary models have two fields: the inflaton and graviton, and the lowest-order parity-violating coupling between them is  dynamical Chern-Simons (dCS) gravity with a decay constant $f$. Here, we show that dCS imprints a parity-violating signal in primordial scalar perturbations. Specifically, we find that, after dCS amplifies one graviton helicity due to a tachyonic instability, the graviton-mediated correlation between two pairs of scalars develops a parity-odd component. This correlation, the primordial scalar trispectrum, is then transferred to the corresponding curvature correlator and thus is imprinted in both large-scale structure (LSS) and the cosmic microwave background (CMB). We find that the parity-odd piece has roughly the same amplitude as its parity-even counterpart, scaled linearly by the degree of gravitational circular polarization $\Pi_{\rm circ} \sim \sqrt{\varepsilon}[H^2/(M_{\rm Pl} f)] \leq 1$, with $\varepsilon$ the slow-roll parameter, $H$ the inflationary Hubble scale, and the upper bound saturated for purely circularly-polarized gravitons. We also find that, in the collapsed limit, the ratio of the two trispectra contains direct information about the graviton's spin. In models beyond standard inflationary dCS, e.g. those with multiple scalar fields or superluminal scalar sound speed, there can be a large enhancement factor $F \gtrsim 10^6$ to the trispectrum. We find that an LSS survey that contains $N_{\rm modes}$ linear modes would place an $n\sigma$ constraint on $\Pi_{\rm circ}r$ of $\sim 0.04\ (n/3)(10^6/F)(10^6/N_{\rm modes})^{1/2}$ from the parity-odd galaxy trispectrum, for tensor-to-scalar ratio $r$. We also forecast for several spectroscopic and 21-cm surveys. This constraint implies that, for high-scale single-field inflation parameters, LSS can probe very large dCS decay constants $f \lesssim 4\times 10^9\ {\rm GeV}(3/n)(F/10^6)\left(N_{\rm modes}/10^6\right)^{1/2}$. Our result is the  first example of a massless particle yielding a parity-odd scalar trispectrum through spin-exchange. 
\end{abstract}

\maketitle

\pagestyle{myheadings}
\markboth{Cyril Creque-Sarbinowski}{Parity-Violating Trispectrum from Chern-Simons Gravity}
\thispagestyle{empty}

\section{Introduction}
Parity (P) violation is maximally realized in the Standard Model through the weak interactions~\cite{Lee:1956qn, Glashow:1961tr}. Moreover, in order to produce the required observed baryon asymmetry of the Universe today, the early Universe is required to violate charge (C) conjugation, its combination CP and Baryon number (B)~\cite{Sakharov:1967dj}. It is therefore not unreasonable to consider scenarios where the Universe violates parity and its source is given by early-Universe dynamics. From an observational standpoint, such parity-violation can be measured in astrophysical systems, large-scale structure (LSS), the cosmic microwave background (CMB), and primordial gravitational waves (pGWs)~\cite{astro-ph/9812088}. In astrophysical systems, parity violation can be potentially be observed through a non-zero correlation of galaxy spins~\cite{1904.01029, 2111.12590} or helical (primordial) magnetic fields~\cite{1410.2250, 1604.06327}. In LSS, the first observable that encapsulates parity violation in the Universe is the galaxy four-point function, as first noted in \cite{2110.12004}, or alternatively its corresponding trispectrum~\cite{2210.02907} (see Refs.~\cite{2206.04227,2206.03625,2210.16320} for applications of these ideas). In addition, LSS parity violation can be probed in any correlator that involves an odd number of galactic shear $B$ modes, such as the galactic shear $E/B$ cross-correlator~\cite{1205.1514, 2001.05930} [or a non-zero curl component]. In the CMB, similar to the galactic shear, parity violation can be measured through correlators that involve an odd number of $B$ modes~\cite{astro-ph/9812088, 1308.6769, 1510.06042}, such as the $E/B$ cross correlator, which can also measure a late-time cosmic birefringence signal~\cite{2011.11254}, as well as both the CMB temperature and polarization trispectra~\cite{1312.5221, 1608.00368}. Finally, chiral pGWs directly encode parity violation in any graviton correlator~\cite{1104.2846, 1106.3228, 1108.0175, 1407.8439, 1706.04627, 2004.00619, 2109.10189}. This vast number of observables then allows us to test a number of early-Universe parity-violating models, most of which are dominated by those that involve inflation~\cite{Guth:1980zm, Freese:1990rb, Martin:2013tda}. Recent hints~\cite{2206.04227, 2206.03625, 2011.11254, 2205.13962} of parity-violation in our Universe have come from the parity-odd galaxy trispectrum and the CMB $E/B$ cross-correlation. 

In the scalar sector, the lowest-order parity-violating correlator of inflation is its trispectrum (assuming statistical isotropy). Ref.~\cite{2210.02907} recently derived a set of no-go theorems that demand a zero parity-odd scalar trispectrum and gave two outs for single-field models. They must either break scale-invariance, via explicit IR-divergences or time-dependent couplings, or obey non-Bunch-Davies initial conditions, one example being ghost inflation~\cite{hep-th/0312100, hep-th/0312099}. Inflationary models with multiple fields have more flexibility and can present parity-breaking signatures in all aforementioned observables~\cite{1203.0302}. Coupling between multiple scalars can break scale-invariance via an IR-divergence ~\cite{2210.02907} and give a parity-odd scalar trispectrum. Coupling between the inflaton and a spin-1 field can break parity via a Chern-Simons term in the presence of a homogeneous $U(1)$ background field~\cite{1312.5221, 1411.2521, 1505.02193, 1608.00368} or with only perturbations if the $U(1)$ field is massive~\cite{1909.01819, 2210.16320}. Chern-Simons gravity~\cite{gr-qc/0308071, 0907.2562} and the Nieh-Yah term in teleparallel gravity~\cite{2007.08038} can generate parity violation in the spin-2 sector.  For strictly massive spin-1 and spin-2 fields, it also possible to construct dimension-5 operators that give large parity-violating signals~\cite{1910.12876, 2004.02887, 2104.08772, 2203.06349, 2208.13790}. Aside from inflation, sources of parity-violation can also come from cosmic strings~\cite{0711.0747, 1110.1718}. The study of parity-violation in inflation is more broadly encapsulated in the Effective Field Theory of Inflation~\cite{0709.0293, 0804.4291}, the Cosmological Collider program~\cite{1503.08043, 1610.06559}, and the Cosmological Bootstrap~\cite{1811.00024, 2203.08121}. Given that the simplest inflationary models contain an inflaton and a graviton, we turn our attention to scalar parity-violation that arises from gravitational interactions, in particular Chern-Simons gravity. 

The lowest-order theory of Chern-Simons gravity that is relevant during inflation is dynamical Chern-Simons (dCS) gravity. Non-dynamical frameworks, where there is no scalar kinetic term, cannot yield inflation, however are useful as a toy model for studying perturbations around black holes~\cite{1110.6241, 1706.08843}. The original method of obtaining dCS gravity is from the low-energy limit of string theory via the Green-Schwartz anomaly cancelling condition~\cite{Green:1987sp}. Additional methods have been investigated, where dCS gravity is induced through Lorentz-violating radiative corrections ~\cite{hep-th/0403205, 0708.3348, 0805.4409}, although these might in fact vanish~\cite{1903.10100}. Moreover, dCS gravity can also arise from spontaneous and dynamical symmetry breaking~\cite{2207.05094, 1602.03191, 1608.08969}. Outside of inflation, dCS gravity can cause alterations to bodies orbiting Earth~\cite{0708.0001},  modifications to gravitational waves from compact binary coalescences~\cite{0904.4501, 1004.4007, 1206.6130, 1705.07924,  2101.11153, 2103.09913, 2208.02805}, and new structures to form around black holes~\cite{2201.02220, 2104.00019}.

 In this paper, we show both that Chern-Simons gravity imprints a parity-breaking signature in the scalar trispectrum and that it can potentially be measured in LSS and the CMB. We bypass the no-go theorem of Ref.~\cite{2210.02907} via the gravitational Chern-Simons term, which changes the graviton mode functions from their standard massless de Sitter form to ones parameterized by a helicity-dependent chiral chemical potential, giving rise to gravitational amplitude-birefringence (circular dichrohism). We find that, in standard Chern-Simons gravity models, the amplitude of the parity-odd signal is roughly the degree of gravitational circular polarization times the tensor-to-scalar ratio. Moreover, we find that, in the collapsed limit, the ratio of the parity-odd trispectrum to the parity-even trispectrum contains direct information about the graviton's spin, allowing one to, in principle, distinguish between different models that yield a parity-odd trispectrum. The appearance of the tensor-to-scalar ratio is due to a consistency condition on the mixed tensor-scalar-scalar bispectrum, yielding a naturally small trispectrum. Moreover,  the degree of circular polarization is saturated at 1 for gravitons of a single helicity. Therefore, to make a detectable signal, a large enhancement factor is required. We show that such an enhancement can come from the presence of multiple light scalar fields or from superluminal scalar sound speeds, however this list is not exhaustive. We point out that, as a result of this enhancement factor, it is possible that a parity-violating trispectrum from gravitational effects can be detected while still having an undetectable tensor-to-scalar ratio. Moreover, models which produce large parity-violating signals could also yield leptogenesis~\cite{hep-th/0403069, 1711.04800}.

This paper is organized as follows. In Section~\ref{sec:th} we present standard dCS gravity in an inflationary context, calculating both the dynamics of the theory and primordial power spectra. We calculate the scalar trispectrum due to graviton-exchange in Section~\ref{sec:in-in} in four settings: one without the dCS coupling, one in standard dCS inflation, and two beyond standard dCS inflation. We relate our calculation of the collapsed limit trispectrum in both standard and beyond-standard dCS models to LSS observables and specific spectroscopic and 21-cm surveys in Section~\ref{sec:obs}. We present a technical discussion in Sec.~\ref{sec:disc} of assumptions and extensions to our calculation and conclude with a general overview in Sec.~\ref{sec:conc}. 

{\it Conventions}: We let $\hbar = c = 1$, use $M_{\rm Pl}^2 = 1/(8\pi G)$ as the reduced Planck mass, and use Fourier conventions
\begin{align}
\nonumber f({\bf x}) = \int \frac{d^3{\bf k}}{(2\pi)^3}e^{-i{\bf k}\cdot{\bf x}}f({\bf k}),\\
\nonumber f({\bf k}) = \int d^3{\bf x}\,e^{i{\bf k}\cdot{\bf x}}f({\bf x}).
\end{align}
When discussing correlators of an observable in Fourier space, we let $\prime$ denote the correlator without volume factors, e.g. $\langle \delta({\bf k})\delta({\bf k}')\rangle = (2\pi)^3\delta_D^{(3)}({\bf k} + {\bf k}')\langle \delta({\bf k})\delta({\bf k}')\rangle^{'}$, with $\delta_D^{(3)}({\bf x})$ the 3D Dirac delta distribution. Moreover, we denote the magnitude of a vector ${\bf k}$ by $k = |{\bf k}|$ and its unit direction by a hat over a bolded letter ${\bf k} = k \hat{\bf k}$. We also use $\prime$ on fields (not on the correlator) to denote derivatives with respect to conformal time, and overdots to denote derivatives with respect to cosmic time. Creation $\hat{b}^\dagger$ and annihilation $\hat{b}$ operators satisfy the commutation relation $[\hat{b}, \hat{b}^\dagger] = (2\pi)^3\delta_D^{(3)}\left({\bf k} - {\bf k}'\right)$.
\section{Standard dCS Inflation}\label{sec:th}
We first review the standard formalism for dynamical Chern-Simons (dCS) gravity during inflation. Specifically, we first present the dynamics of both the inflaton's and graviton's spatial perturbations  during standard dCS inflation and then finish by calculating their respective primordial power spectra. 

The dynamics of standard dCS inflation are obtained through its vacuum action $S$, 
\begin{align}\label{eq:dCS}
S &= \int d^4x \sqrt{-g}\left[- \frac{1}{2}g^{\mu\nu}\left(\partial_\mu \phi\right)\left(\partial_{\nu}\phi\right) + V(\phi) + \frac{M_{\rm Pl}^2}{2}R - \frac{\phi}{4f}{^*}RR \right],
\end{align}
with $g_{\mu\nu}$ the spacetime metric and $g$ its determinant, $\phi$ the inflaton and $V(\phi)$ its potential, $R$ the Ricci scalar, and $f$ the dCS decay constant for the process $\phi \rightarrow hh$. Since the potential $V(\phi)$ can be somewhat arbitrary as long as it admits a slow-roll solution and since it does not enter our calculations at lowest order in slow roll, we will leave it unspecified. However, in order to obtain a sense of scales, we note that, during slow roll, the potential is roughly constant, $V \sim \Lambda^4$, and so is related to the Hubble scale $H$ during inflation, through the Friedmann equations, by $\Lambda \sim \sqrt{H M_{\rm Pl}} = 10^{16}\ {\rm GeV}\ [H/(10^{14}\ {\rm GeV})]^{1/2}$. The Pontryagin density is defined as
\begin{align}
{^*}RR &= {^*}\tensor{R}{^\rho_\sigma^\mu^\nu} R^\sigma_{\rho\mu\nu}
\end{align}
where
\begin{align}
{^*}\tensor{R}{^\rho_\sigma^\mu^\nu} &= \frac{1}{2}\epsilon^{\mu\nu\alpha\beta}R^\rho_{\sigma\alpha\beta}
\end{align}
is the Hodge dual of the Riemann tensor and $\epsilon^{\mu\nu\alpha\beta}$ is the Levi-Civita tensor. We work in a perturbed flat Friedmann-Lem\^aitre-Robertson-Walker (FLRW) spacetime using conformal time $\tau$ and parameterized by both the scale factor $a(\tau) = -1/(H\tau)$ and tensor (graviton) perturbations $h_{ij}$,
\begin{align}
g_{\mu\nu} &= a^2(\tau)\left(\begin{array}{cc}
-1 & 0\\
0 & \delta_{ij} + h_{ij}\\
\end{array}\right),
\end{align} 
so that the graviton is both transverse, $\partial^ih_{ij} = 0$, and traceless, $h^i_i = 0$, and the determinant of the metric is $\sqrt{-g} = a^4(\tau)$.  

The background (i.e. spatially-independent) portion of the action in Eq.~\eqref{eq:dCS} helps give rise to inflation itself. The dCS action at first order in all perturbations must vanish since a) the background inflaton satisfies its equations of motions and b) a transverse and traceless tensor cannot form a scalar. Therefore, the lowest-order corrections to the background evolution come in at second order in scalar and tensor perturbations, when the Pontryagin density takes the form~\cite{hep-th/0401153}
\begin{align}
{^*}RR = \frac{2\tilde{\epsilon}^{ijk}}{a^4(\tau)}\left[h''_{il}\left(\partial_jh^l_k\right)' + \left(\partial_m h'_{li}\right)\left(\partial_j \partial^l h_{k}^m + \partial_k \partial^m  h_{j}^l\right)\right],
\end{align}
with $\tilde{\epsilon}^{ijk}$ the Levi-Civita symbol, related to its tensor equivalent by $\epsilon^{0ijk} = \tilde{\epsilon}^{ijk}/\sqrt{-g}$. The second-order action for dCS is then
\begin{align}
S^{(2)} &= \int d^4x\left\{-\frac{1}{2}a^2\left(\delta\phi'\right)^2 - \frac{1}{2}a^2\left(\partial_i \delta\phi\right)\left(\partial^i\delta\phi\right) + \frac{M_{\rm Pl}^2}{8}a^2\left[\left(h^i_j\right)'\left(h^j_i\right)' + \left(\partial_k h^i_j\right)\left(\partial^k h^j_i\right)\right] - a^4\frac{\phi(\tau)}{4f}{^*}RR\right\},
\end{align} 
where we give $\phi$ an explicit argument in the dCS term to emphasize it is only the background piece of the inflaton.

We expand each field in terms of its 3-dimensional Fourier modes ${\bf k}$ and the graviton in terms of its helicities $h \in \{+, -\}$, 
\begin{align}
\delta\phi(\tau, {\bf x}) &= \int \frac{d^3{\bf k}}{(2\pi)^3}\left[u(\tau, {\bf k}) b_0({\bf k}) + u^*(\tau, -{\bf k})\hat{b}_0^\dagger(-{\bf k})\right]e^{i{\bf k}\cdot {\bf x}},\\
\nonumber h_{ij}(\tau, {\bf x}) &= \sum_{h = \pm}\int\frac{d^3 {\bf k}}{(2\pi)^3}[u_h(\tau, {\bf k})\hat{b}_h({\bf k})e_{ij}^h({\bf k}) \\
&\hspace{2cm}+ u_h^*(\tau, -{\bf k})\hat{b}^\dagger_h(-{\bf k})e_{ij}^{h*}(-{\bf k})]e^{i{\bf k}\cdot {\bf x}}\label{eq:grav_mode}.
\end{align}

As a result, the mode functions obey the equations of motion~\cite{hep-ph/9907244, hep-th/0403069, hep-th/0501153, 0806.4594, 2204.12874, 1607.03735}
\begin{align}
u''(\tau, {\bf k}) -\frac{2}{\tau}u'(\tau, {\bf k}) + k^2 u(\tau, {\bf k}) &= 0,\\
\label{eq:g_EOM}u_\pm''(\tau, {\bf k}) - \frac{2}{\tau}u_\pm'(\tau, {\bf k}) +  \left(k^2 \mp \frac{2 k \mu}{\tau}\right)u_\pm(\tau, {\bf k}) &= 0,
\end{align}
where $\mu = \sqrt{\varepsilon/2}\left(M/f\right)$ is the spin-2 chiral chemical potential in Hubble units with $\varepsilon = (\dot{\phi}^2/2)/(M_{\rm Pl} H)^2$ the standard slow-roll parameter and $M = \left(H/M_{\rm Pl}\right)^2 M_{\rm Pl}$ a mass scale.~\footnote{If there are fermions in the theory, then by the chiral anomaly~\cite{Delbourgo:1972xb}, $\partial_\mu J^\mu_5=  {^*}RR/(384\pi^2)$, with $J^\mu_5$ the axial current. Since $J^0_5 = n_L - n_R$, where $n_\alpha$ is the number density of $\alpha$-handed fermions, then after inserting the anomaly into the  action, integrating by parts, and canonically normalizing our fields, the dCS term at lowest order becomes $\sim H\mu(n_L - n_R)$. This term allows us to identify $\mu$ as a chemical potential that induces a difference of left-handed and right-handed fermions, i.e. the chemical potential induces net chirality. In addition, we point out that, in dCS studies $\mu$ is often called $\Theta$~\cite{hep-th/0403069, hep-th/0501153, hep-th/0410230} or the Chern-Simons mass $M_{\rm CS}/(2H)$. For readers more familiar with $U(1)$ Chern-Simons theories, the analogous spin-1 chemical potential is often called $\xi$~\cite{0908.4089, 1212.1693, 2002.02952, 2110.10695, 2204.12874}. 
} In order to obtain the graviton equations of motion in the form of Eq.~\eqref{eq:g_EOM}, we assumed a small chemical potential, $\mu \lesssim 1$ (or more precisely $k\tau\mu\lesssim 1$ for one graviton helicity), so that graviton perturbations that originate within the Hubble horizon $k\tau \gtrsim 1$ have physical modes that are not ghost-like in nature~\cite{0806.4594, 1208.4871, 1706.04627}. As we will see, we also require $\mu \lesssim 1$ so as not to overproduce tensor modes through Eq.~\eqref{eq:r}. As emphasized by the potential term in Eq.~\eqref{eq:g_EOM}, one graviton helicity develops a tachyonic instability for $k\tau < 2\mu \lesssim 1$. Hence, for one graviton helicity, sub-horizon modes go ghost, while super-horizon modes go tachyonic, the latter processes yielding gravitational amplitude-birefringence. We refer the reader to Ref.~\cite{1406.4550} for a nice review on the difference between ghosts and tachyons. 

While the equations of motion as we have written them above are useful in analyzing their sub- and super-horizon limiting behavior, they are not written in canonical form (i.e. of a harmonic oscillator with possibly time-dependent frequency). Their canonical forms are
\begin{align}
\left[a(\tau)u(\tau, {\bf k})\right]'' + \left(k^2 - \frac{2}{\tau^2}\right)\left[a(\tau)u(\tau, {\bf k})\right] &= 0,\\
\left[a(\tau)u_\pm(\tau, {\bf k})\right]'' + \left(k^2 - \frac{2}{\tau^2} \mp \frac{2 k \mu}{\tau}\right)\left[a(\tau)u_\pm(\tau, {\bf k})\right] &= 0.
\end{align}
Solving for the canonically-normalized mode functions $a(\tau)u(\tau, {\bf k})$ and $a(\tau)u_\pm(\tau, {\bf k})$ and imposing both canonical quantization and vacuum (Bunch-Davies) initial conditions on them, we obtain
\begin{align}
\label{eq:0_mode}u(\tau, {\bf k}) &= \frac{H}{\sqrt{2k^3}}(1 + ik\tau)e^{-ik\tau}\\
\label{eq:g_mode}u_\pm(\tau, {\bf k})&= \frac{H}{M_{\rm Pl}\sqrt{k^3}}(1 + ik\tau)e^{-ik\tau}\left[\frac{U\left(2 \mp i\mu, 4, 2 i k \tau\right)}{U\left(2, 4, 2ik\tau\right)}\right]\exp\left(\pm \frac{\pi}{2}\mu\right), 
\end{align}
where $U$ is the confluent hypergeometric function of the second kind. In the first two arguments of $U$, the numbers 2 and 4 correspond to the spin $s = 2$ and twice the spin of the graviton~\cite{2208.13790}. The factor of $\pi/2$ in the exponent is a non-linear realization of a logarithmic IR divergence in the standard graviton mode functions through the Cosmological Optical Theorem~\cite{2009.02898}.~\footnote{We thank David Stefanyszyn for pointing this fact out.} The polarization vectors  $e_i^h({\bf k})$ are normalized so that $e_i^h({\bf k})e_i^{h'*}\left({\bf k}\right) = \delta_{h h'}$ and the polarization matrices $e_{ij}^h({\bf k})$ are normalized so that $e_{ij}^h({\bf k})e_{ij}^{h'*}({\bf k})= 2\delta_{h h'}$. Eq.~\eqref{eq:0_mode} and Eq.~\eqref{eq:g_mode} specify the dynamics of dCS inflation. 

The inflaton perturbations $\delta\phi$ are related to the conserved comoving primordial curvature perturbation $\zeta$ through the spatially-flat gauge relation at linear order $\zeta({\bf k}) = -H[\delta\phi({\bf k})/\dot{\phi}] = -\delta\phi({\bf k})/(\sqrt{2\varepsilon} M_{\rm Pl})$. Using this relation and the previous mode functions, the scalar and (polarized) tensor primordial power spectra on super-horizon scales, $k\tau \ll 1$, are then 
\begin{align}\label{eq:PS}
P_\zeta(k) &= \frac{|u(\tau, {\bf k})|^2}{2\varepsilon M_{\rm Pl}^2}\Bigg{|}_{k\tau \ll 1} = \frac{1}{4\varepsilon k^3}\left(\frac{H}{M_{\rm Pl}}\right)^2 = \frac{2\pi^2 A_s}{k^3}\\
P_\pm(k) &=  |u_\pm(\tau, {\bf k}) e^\pm_{ij}({\bf k})|^2 \Big{|}_{k\tau \ll 1} = \frac{\Theta_{\pm}}{2}\bar{P}_t(k)\label{eq:PS_g},
\end{align}
where $A_s = 2.1\times 10^{-9}$ is the amplitude of the primordial scalar power spectrum~\cite{1807.06209}, $\Theta_\pm = \exp\left(\pm \pi \mu\right)/\left[\Gamma\left(2 - i \mu\right)\Gamma\left(2 + i\mu\right)\right] = 1 \pm \pi \mu + O\left(\mu^2\right)$ are the enhancement/suppression factors for the $\pm$ graviton modes and 
$\bar{P}_t(k) = \bar{A}_t k^{-3}$ the standard graviton (tensor) primordial power spectrum with primordial tensor amplitude $\bar{A}_t  = 4(H/M_{\rm Pl})^2$. As emphasized by the exponential in Eq.~\eqref{eq:g_mode}, one graviton polarization is enhanced and the other suppressed through the tachyonic instability in its equation of motion.  As a result, the tensor-to-scalar ratio in dCS inflation, $r = [P_+(k) + P_-(k)]/P_\zeta(k)$, is enhanced relative to the standard (non-dCS) ratio, $\bar{r} = \bar{P}_t(k)/P_\zeta(k) = 16\varepsilon$, by a factor
\begin{align}\label{eq:r}
r = \Theta_{\rm even}\bar{r},
\end{align}
with $\Theta_{\rm even} = \left(\Theta_+ + \Theta_-\right)/2 = \cosh\left(\pi \mu\right)/\left[\Gamma\left(2 - i \mu\right)\Gamma\left(2 + i \mu\right)\right] = 1 + (2\pi^2 - 3)\mu^2/3 + O\left(\mu^4\right)$.~\footnote{This factor is different than Refs.~\cite{0806.4594, 1706.04627} by the additional factors of $\Gamma(2 \pm i\mu)$ that come from the asymptotic expansion of the hypergeometric function $U$. However, they only make a $\sim 10\%$ difference at second order in $\mu$.} For later use, we also define the antisymmetric combination $\Theta_{\rm odd} = \left(\Theta_+ - \Theta_-\right)/2 = \sinh\left(\pi \mu\right)/\left[\Gamma\left(2 - i \mu\right)\Gamma\left(2 + i \mu\right)\right]= \pi \mu + O\left(\mu^2\right)$ and a measure for the degree of gravitational circular polarization $ \Pi_{\rm circ} = [P_+(k) - P_-(k)]/[P_+(k) + P_-(k)] = \Theta_{\rm odd}/\Theta_{\rm even}$. Quantitatively, the degree of gravitational circular polarization takes values according to
\begin{align}\label{eq:g_circ}
\Pi_{\rm circ} = 0.9\left(\frac{\varepsilon}{10^{-2}}\right)^{1/2}\left(\frac{H}{10^{14}\ {\rm GeV}}\right)^2\left(\frac{10^{9}\ {\rm GeV}}{f}\right) + O\left(\mu^2\right),
\end{align}
 with saturation occurring at $\Pi_{\rm circ} = \pm 1$ when only the $\pm$ graviton helicity is populated (i.e. for purely circularized gravitons). Given the dCS-inflation dynamics, primordial power spectra,  and degree of gravitational circular polarization, Eqs.~(\ref{eq:0_mode}- \ref{eq:g_circ}), we now move to our main calculation: the connected primordial scalar trispectrum due to graviton-exchange.    

\section{Graviton-Mediated Primordial Scalar Trispectrum}\label{sec:in-in}
\begin{figure}[ht]
\begin{tikzpicture}
\begin{feynman}
\vertex (a1) {\(\tau = 0\)};
\vertex [right=of a1] (a2) [label=above:$\delta\phi$] ;
\vertex [right=of a2] (a3) ; 
\vertex [right=of a3] (a4) [label=above:$\delta\phi$];
\vertex [right=of a4] (a5) [label=above:$\delta\phi$];
\vertex [right=of a5] (a6);
\vertex [right=of a6] (a7) [label=above:$\delta\phi$];
\vertex [right=of a7] (a8);
\node [below=of a3, crossed dot] (v1);
\node [below=of a6, crossed dot] (v2);
\vertex [below=of a1] (b1) {\(\tau < 0\)};
\diagram*{
(a1) -- [plain, very thick] (a2) -- [plain, very thick] (a3) -- [plain, very thick] (a4) -- [plain, very thick] (a5) -- [plain, very thick] (a6) -- [plain, very thick] (a7)-- [plain, very thick] (a8),
(a2) -- [plain] (v1),
(a4) -- [plain] (v1),
(a5) -- [plain] (v2),
(a7) -- [plain] (v2),
(v1) -- [graviton, half right, edge label = \(h\)] (v2)
};
\end{feynman}
\end{tikzpicture}
\caption{Diagram of the scalar trispectrum mediated by graviton exchange. The external legs represent scalar propagators, while the wavy line represents the graviton propagator. The thick horizontal line is the conformal boundary $\tau = 0$, so that the propagators live in the bulk $\tau < 0$. Each vertex comes from the standard kinetic term $\bigotimes = (1/2)a^2 h^{ij}\partial_i \delta\phi \partial_j \delta\phi$. In general, such a diagram is called a spin-exchange diagram. }\label{fig:4PTGE}
\end{figure}
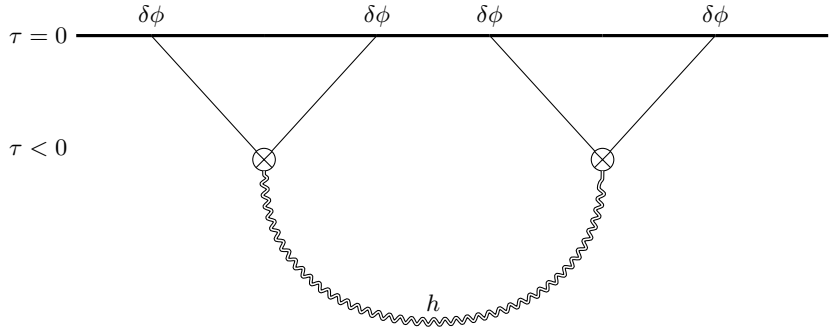

We present our calculation of the connected ($c$) primordial scalar trispectrum due to graviton exchange (GE) in inflationary dCS,  $\langle \delta \phi({\bf k}_1)\delta\phi({\bf k}_2)\delta\phi({\bf k}_3)\delta\phi({\bf k}_4)\rangle^{\rm GE}_c$, which we diagrammatically show in Fig.~\ref{fig:4PTGE}. For simplicity, we often refer to this trispectrum as the GE scalar trispectrum. We perform this calculation using the in-in formalism~\cite{astro-ph/0210603, hep-th/0506236, 1703.10166}, specifically using the Schwinger-Keldysh diagrammatic rules of Ref.~\cite{1703.10166}. Using these rules, we first calculate the GE scalar trispectrum without specifying the functional form of the graviton mode functions in Sec.~\ref{subsec:gen_calc}. This lack of specification then allows us to both review the results for this trispectrum without dCS in Sec.~\ref{subsec:wo_dCS} and calculate our new results with standard dCS in Sec.~\ref{subsec:w_dCS}. Moreover, in both Sec.~\ref{subsec:wo_dCS} and Sec.~\ref{subsec:w_dCS}, we present the full GE trispectrum and its collapsed limit. We finish our calculation in Sec.~\ref{subsec:b_dCS} by extending the standard calculation, in the collapsed limit, to two models beyond standard inflationary dCS. 

Three points are necessary before the calculation. First, the diagram in Fig.~\ref{fig:4PTGE} exists in inflation without the gravitational Chern-Simons coupling, as the graviton-scalar-scalar coupling comes from the standard scalar kinetic term
\begin{align} \label{eq:kin}
\frac{1}{2}\sqrt{-g}g^{\mu\nu}\partial_\mu \delta\phi\partial_\nu\delta\phi = \frac{1}{2}a^2\eta^{\mu\nu}\partial_\mu\delta\phi\partial_\nu\delta\phi + \frac{1}{2}a^2h^{ij}\partial_i\delta\phi\partial_j\delta\phi,
\end{align}
with $\eta^{\mu\nu} = {\rm diag}(-1, 1, 1, 1)$ the Minkowski metric, and its value was calculated in the setting of single-field inflation in Refs.~\cite{0811.3934, 2212.07370}. The effect of dCS on this diagram is only to change the graviton mode functions by the additional factors in Eq.~\eqref{eq:g_mode}. Alternative graviton-scalar-scalar couplings that are the same, or lower, order in derivatives vanish due to the symmetric and transverse-traceless nature of the graviton. Higher-derivative couplings are parameterized by the form $\left[c_n/\Lambda_{(n)}^{2n}\right]h^{ij}\left[\partial_i\left(\partial_{\alpha_1}\ldots\partial_{\alpha_n}\right)\delta\phi\right]\left[\partial_j \left(\partial^{\alpha_1}\ldots \partial^{\alpha_n}\right)\delta\phi\right]$ with $c_n$ a dimensionless coupling constant and $\Lambda_{(n)}$ some cutoff scale. These higher-derivative terms are then suppressed relative to the standard kinetic interaction by factors of the inflaton momenta divided by the cutoff.

Second, reality of the real-space four-point function, $\langle\delta\phi({\bf x}_1)\delta\phi({\bf x}_2)\delta\phi({\bf x}_3)\delta\phi({\bf x}_4)\rangle_c$, imposes that the parity-even component of the full scalar trispectrum is exactly the real part of the full scalar trispectrum, $\langle \delta \phi({\bf k}_1)\delta\phi({\bf k}_2)\delta\phi({\bf k}_3)\delta\phi({\bf k}_4)\rangle^{{\rm even}}_c = {\rm Re}\langle \delta \phi({\bf k}_1)\delta\phi({\bf k}_2)\delta\phi({\bf k}_3)\delta\phi({\bf k}_4)\rangle_c$. The same logic imposes an analogous relation between the parity-odd component and the imaginary part,  $\langle \delta \phi({\bf k}_1)\delta\phi({\bf k}_2)\delta\phi({\bf k}_3)\delta\phi({\bf k}_4)\rangle^{{\rm odd}}_c = {\rm Im}\langle \delta \phi({\bf k}_1)\delta\phi({\bf k}_2)\delta\phi({\bf k}_3)\delta\phi({\bf k}_4)\rangle_c$. 

Third, in inflationary dCS, the lowest-order contribution to the full parity-odd scalar trispectrum is in fact the diagram of two pairs of scalars Wick-contracted through graviton exchange, ${\rm Im}\langle \delta\phi({\bf k}_1)\delta\phi({\bf k}_2)\delta\phi({\bf k}_3)\delta\phi({\bf k}_4)\rangle_c\approx{\rm Im}\langle \delta\phi({\bf k}_1)\delta\phi({\bf k}_2)\delta\phi({\bf k}_3)\delta\phi({\bf k}_4)\rangle_c^{\rm GE}$. Hence, our calculation is the dominant primordial parity-odd signal. 

\subsection{General Calculation}\label{subsec:gen_calc}
\begin{figure}[ht]

\begin{tikzpicture}
\begin{feynman}[small]
\vertex (a1);
\vertex [right=of a1] (a2) [label=above:$\delta\phi$] ;
\vertex [right=of a2] (a3) ; 
\vertex [right=of a3] (a4) [label=above:$\delta\phi$];
\vertex [right=of a4] (a5) [label=above:$\delta\phi$];
\vertex [right=of a5] (a6);
\vertex [right=of a6] (a7) [label=above:$\delta\phi$];
\vertex [right=of a7] (a8);
\vertex [below=of a3] (v1);
\vertex [below=of a6] (v2);
\vertex [right=of a8] (a9);
\node [node font=\Large, below=of a9] (b1) {\(\ \ \pm\ \ \ \)};
\diagram*{
(a1) -- [plain, very thick] (a2) -- [plain, very thick] (a3) -- [plain, very thick] (a4) -- [plain, very thick] (a5) -- [plain, very thick] (a6) -- [plain, very thick] (a7)-- [plain, very thick] (a8),
(a2) -- [plain] (v1),
(a4) -- [plain] (v1),
(a5) -- [plain] (v2),
(a7) -- [plain] (v2),
(v1) -- [graviton, half right, edge label = \(h_+\)] (v2),
};
\end{feynman}
\end{tikzpicture}
\begin{tikzpicture}
\begin{feynman}[small]
\vertex (a9);
\vertex [right=of a9] (a10);
\vertex [right=of a10] (a11) [label=above:$\delta\phi$] ;
\vertex [right=of a11] (a12) ; 
\vertex [right=of a12] (a13) [label=above:$\delta\phi$];
\vertex [right=of a13] (a14) [label=above:$\delta\phi$];
\vertex [right=of a14] (a15);
\vertex [right=of a15] (a16) [label=above:$\delta\phi$];
\vertex [right=of a16] (a17);
\vertex [below=of a12] (v3);
\vertex [below=of a15] (v4);
\diagram*{
(a10) -- [plain, very thick] (a11) -- [plain, very thick] (a12) -- [plain, very thick] (a13) -- [plain, very thick] (a14) -- [plain, very thick] (a15) -- [plain, very thick] (a16)-- [plain, very thick] (a17),
(a11) -- [plain] (v3),
(a13) -- [plain] (v3),
(a14) -- [plain] (v4),
(a16) -- [plain] (v4),
(v3) -- [graviton, half right, edge label = \(h_-\)] (v4)
};
\end{feynman}
\end{tikzpicture}
\caption{Visual representation for parity-even and -odd  trispectra corresponding to Eq.~\eqref{eq:Re_4} and Eq.~\eqref{eq:Im_4}, respectively. The sum of both graviton helicities $h_\pm$ corresponds to the parity-even component and the difference to the parity-odd component.  We leave the value of the vertices unspecified as these drawings are not computational diagrams. Otherwise, we use a similar labeling scheme as Fig.~\ref{fig:4PTGE}}\label{fig:ParityGE}
\end{figure}
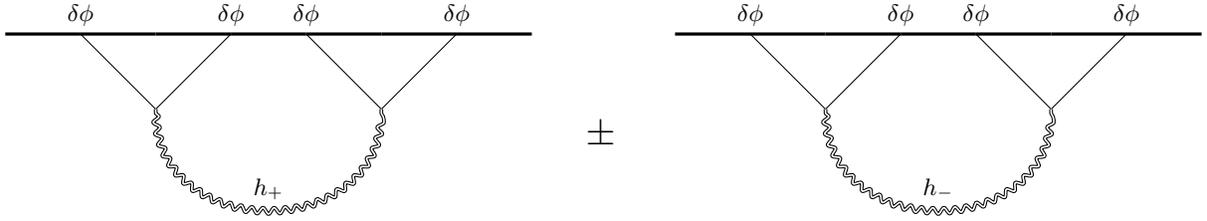

We compute both the lowest-order contribution to the parity-odd trispectrum and the dCS corrections to the graviton-mediated parity-even trispectrum. In order to perform this calculation, we follow the Schwinger-Keldysh (SK) diagrammatic rules presented in Ref.~\cite{1703.10166}. Roughly speaking, and adapted to our specific calculation, they say the following:
\begin{enumerate}
\item[Step 1.]{Label each external leg in Fig.~\ref{fig:4PTGE} by one of four momenta ${\bf k}_1$, ${\bf k}_2$, ${\bf k}_3$, ${\bf k}_4$ and consider the set of all distinct (in the sense that they obey the symmetries of the diagram) labelings. There are three such diagrams in this set and each diagram we refer to as a channel.  Explicitly, let ${\bf K}_{s} = \{{\bf k}_1, {\bf k}_2, {\bf k}_3, {\bf k}_4\},\ {\bf K}_t = \{{\bf k}_1, {\bf k}_3, {\bf k}_2, {\bf k}_4\},$ and ${\bf K}_u = \{{\bf k}_1, {\bf k}_4, {\bf k}_3, {\bf k}_2\}$ be three ordered lists of vectors, each corresponding to one of these channels. Due to translation invariance, each channel obeys the relations ${\bf K}_I^1 + {\bf K}_I^2 = - {\bf K}_I^3 - {\bf K}_I^4 = {\bf k}_I$, $I \in \{s, t, u\}$, where ${\bf k}_I$ is a Mandelstam variable (e.g. ${\bf k}_s = {\bf k}_1 + {\bf k}_2$). }
\item[Step 2.]{For each channel, label each vertex with a $\pm$. Different vertices do not need to have the same sign. That is, there will be four combinations, $++, +-, -+,$ and $--$, for every channel. For each combination, write down a definite integral over all conformal time and one factor of the vertex coupling constant $(1/2)a^2$,   $(1/4)\int [d\tau_1/(H\tau_1)^2][d\tau_2/(H\tau_2)^2]$. For vertices with an odd number of $-$'s, multiply by a factor of $-1$. To be clear, the labels $\pm$ do not represent helicities , but rather have to do with the SK formalism (see Ref.~\cite{1703.10166} for more details).}
\item[Step 3.]{For each external leg connected to a vertex, write, within the integrals, one factor of the scalar bulk-to-boundary propagator
\begin{align}\label{eq:SBBP}
\mathcal{G}_\pm({\bf k}, \tau) &= \frac{H^2}{2k^3}(1 \mp ik\tau)e^{\pm ik\tau},
\end{align}
with each factor having same sign as the associated vertex.  The scalar propagator has momentum equal to the momentum of the boundary scalar perturbation of the external leg and is evaluated at the conformal time of the associated vertex. Here, boundary refers to the conformal time boundary $\tau = 0$ and bulk to $\tau < 0$. }
\item[Step 4.]{For each vertex combination, multiply by the graviton-helicity-dependent (but time-independent) vertex  coupling $\mathcal{P}_h({\bf K}_I)= \left[e^h_{a b}({\bf k}_I)\left(K_I^1\right)^a \left(K_I^2\right)^b\right]\left[e^{h*}_{cd}({\bf k}_I)\left(K_I^3\right)^c\left(K_I^4\right)^d\right]$. We call this factor the polarization portion later. Note that $\mathcal{P}_-({\bf K}_I) = \mathcal{P}^*_+({\bf K}_I)$. }
\item[Step 5.]{
In addition, write down one of the corresponding bulk propagators for a graviton of helicity $h$,
\begin{align}
++ &\quad u_h\left[{\bf k}_I, {\rm max}\left(\tau_1, \tau_2\right)\right]u_h^*\left[{\bf k}_I, {\rm min}\left(\tau_1, \tau_2\right)\right] ,\\
+- &\quad u_h^*({\bf k}_I, \tau_1)u_h({\bf k}_I, \tau_2),\\
-+ &\quad u_h({\bf k}_I, \tau_1)u_h^*({\bf k}_I, \tau_2),\\
-- &\quad u_h\left[{\bf k}_I, {\rm min}\left(\tau_1, \tau_2\right)\right]u_h^*\left[{\bf k}_I, {\rm max}\left(\tau_1, \tau_2\right)\right].
\end{align}
Note that, compared to usual expressions for the bulk propagator, we have traded the Heaviside function for ${\min}$ and ${\max}$ functions to achieve a more compact notation.
}

\item[Step 6.]{Sum over all channels, graviton helicities, and vertex combinations to obtain the final result.}
\end{enumerate}
We split the calculation of the trispectrum according to the above rules into two pieces: the polarization portion, which is time-independent, and the mode-function integration, which is time-dependent. Explicitly, and after some simplification, we write down the scalar trispectrum in the compactified form 
\begin{align}\label{eq:compact}
\langle \delta\phi({\bf k}_1)\delta\phi({\bf k}_2)\delta\phi({\bf k}_3)\delta\phi({\bf k}_4)\rangle_c^{\prime{\rm GE}} &= -\frac{1}{2}\sum_{I, h} \mathcal{P}_h({\bf K}_I){\rm Tvar}_h({\bf K}_I),
\end{align}
with
\begin{align}\label{eq:tvar}
{\rm Tvar}_h({\bf K}_I) &= {\rm Re}\left[\mathcal{I}^{(2)}_h({\bf K}_I) - \mathcal{I}^{(1)}_h({\bf K}^1_I, {\bf K}_I^2)\bar{\mathcal{I}}^{(1)}_h({\bf K}_I^3, {\bf K}_I^4)\right]\\
\label{eq:I2}\mathcal{I}^{(2)}_h({\bf K}_I) &= \int_{-\infty}^0 \int_{-\infty}^{0} \frac{d\tau_1}{(H\tau_1)^2} \frac{d\tau_2}{(H\tau_2)^2} \mathcal{G}_+({\bf K}_I^1, \tau_1)\mathcal{G}_+({\bf K}_I^2, \tau_1)\mathcal{G}_+({\bf K}_I^3, \tau_2)\mathcal{G}_+({\bf K}_I^4, \tau_2)u_h[{\bf k}_I, {\rm max}(\tau_1, \tau_2)]u_h^*[{\bf k}_I, {\rm min}(\tau_1, \tau_2)]\\ 
\label{eq:I1}\mathcal{I}^{(1)}_h({\bf k}_i, {\bf k}_j) &= \int_{-\infty}^0\frac{d\tau}{(H\tau)^2}\ \mathcal{G}_+({\bf k}_i, \tau)\mathcal{G}_+({\bf k}_j, \tau) u_h^*({\bf k}_i + {\bf k}_j, \tau),
\end{align} 
the time-ordered variance, Tvar, of the tensor-scalar-scalar bispectrum $\langle \delta\phi \delta\phi h_{ij}\rangle_h$~\cite{1407.8204, 1504.05993}. We give Eq.~\eqref{eq:tvar} this name as it is the difference between the unfactorizable product of two temporally-connected bispectrum, Eq.~\eqref{eq:I2}, and a factorized product of two temporally-disconnected bispectra, Eq.~\eqref{eq:I1}. In fact, in the large-scale limit ${\bf k}_I \ll {\bf K}_I^i, i \in \{1, 2, 3, 4\}$, where the the propagator momentum can be taken to be small, ${\bf k}_I \tau \ll 1$, the time-ordered variance is exactly the factorized product of two tensor-scalar-scalar bispectra in the same limit,
\begin{align}\label{eq:2bi}
{\rm Tvar}_h\left({\bf K}_I\right)\Big{|}_{{\bf k}_I \ll {\bf K}_I^i} &= {\rm Re}\left[\int_{-\infty}^0 \frac{d\tau}{(H\tau)^2} \mathcal{G}_+({\bf K}_I^1, \tau) \mathcal{G}_+\left({\bf K}_I^2, \tau\right)u_h\left({\bf k}_I, 0\right)\right]{\rm Re}\left[\int_{-\infty}^0\frac{d\tau}{(H\tau)^2} \mathcal{G}_+({\bf K}_I^3, \tau)\mathcal{G}_+\left({\bf K}_I^4, \tau\right)u^*_h({\bf k}_I, 0)\right],
\end{align} 
which emphasizes that this structure is representative of the underlying cutting rules applied to Fig.~\ref{fig:4PTGE}~\cite{2112.03448}. We point out that the overbar in $\bar{\mathcal{I}}_h^{(1)}$ indicates complex conjugation. From Eq.~\eqref{eq:compact} and the noted polarization identity in Step 4, it follows that the even and odd scalar trispectra are a sum and difference of the time-ordered variance due to each helicity summed over all channels,
\begin{align}\label{eq:Re_4}
{\rm Re}\langle \delta\phi({\bf k}_1)\delta\phi({\bf k}_2)\delta\phi({\bf k}_3)\delta\phi({\bf k}_4)\rangle_c^{\prime {\rm GE}} &= -\frac{1}{2}\sum_I{\rm Re}\left[\mathcal{P}_+({\bf K}_I)\right]\left[{\rm Tvar}_+\left({\bf K}_I\right) + {\rm Tvar}_-\left({\bf K}_I\right)\right],\\ \label{eq:Im_4}
{\rm Im}\langle \delta\phi({\bf k}_1)\delta\phi({\bf k}_2)\delta\phi({\bf k}_3)\delta\phi({\bf k}_4)\rangle_c^{\prime {\rm GE}} &= -\frac{1}{2}\sum_I {\rm Im}\left[\mathcal{P}_+({\bf K}_I)\right]\left[{\rm Tvar}_+\left({\bf K}_I\right) - {\rm Tvar}_-\left({\bf K}_I\right)\right], 
\end{align}
which we visually represent in Fig.~\ref{fig:ParityGE}.
\subsubsection{Polarization Portion}
Beginning with polarization, we decompose the spin-2 polarization matrices into a product of spin-1 polarization vectors of the same helicity $e^h_{ij}({\bf k}) = \sqrt{2}e^h_i({\bf k})e^h_j({\bf k})$~\cite{1012.1079}. Then, we use the orthogonality and completeness of the spin-1 polarization vectors over the 2D complex plane, $e^h_i({\bf k})e^{h*}_j({\bf k}) = (1/2)(\delta_{ij} - \hat{k}_i\hat{k}_j -ih\tilde{\epsilon}_{ijl}\hat{k}^l)$~\cite{1505.02193}, to obtain  
\begin{align}\label{eq:P_vec}
\mathcal{P}_h\left({\bf K}_I\right)  = \frac{1}{2}&\left[{\bf K}_I^1\cdot{\bf K}_I^3 - ({\bf K}_I^1\cdot\hat{\bf k}_{I})({\bf K}_I^3\cdot\hat{\bf k}_{I}) - ih{\bf K}_I^1 \cdot({\bf K}_I^3 \times \hat{\bf k}_{I})\right]^2
\end{align}
where we used the identity ${\bf K}_I^1 + {\bf K}_I^2 = -{\bf K}_I^3 - {\bf K}_I^4 = {\bf k}_I$ to substitute for ${\bf K}_I^2$ and ${\bf K}_I^4$. In addition, we set up a (channel-dependent) coordinate system using the three rotated orthonormal basis vectors $\hat{\bf e}_1 = R_I \hat{\bf x}, \hat{\bf e}_2 = R_I \hat{\bf y},$ and  $\hat{\bf e}_3 = R_I \hat{\bf z}$, with $\hat{\bf x}\ (\hat{\bf y})\ [\hat{\bf z}]$ , the unit vector in the $x\ (y)\ [z]$  direction and $R_I$ the $SO(3)$ matrix to rotate $\hat{\bf z}$ into $\hat{\bf k}_I$ so that $\hat{\bf e}_3 = \hat{\bf k}_I$. With this basis, we then write each of the individual momentum vectors in spherical coordinates, ${\bf K}_I^i =  K_I^i\left[\sin\left(\theta_I^i\right)\cos\left(\varphi_I^i\right)\hat{\bf e}_1 + \sin\left(\theta_I^i\right)\sin\left(\varphi_I^i\right)\hat{\bf e}_2 + \cos\left(\theta_I^i\right)\hat{\bf e}_3\right]$. Therefore, the polarization portion, Eq.~\eqref{eq:P_vec}, is 
\begin{align}
\mathcal{P}_h\left({\bf K}_I\right) &= \frac{1}{2}\left[K_I^1 K_I^3 \sin\left(\theta_I^1\right)\sin\left(\theta_I^3\right)\right]^2 e^{2ih \chi_I},
\end{align}
with $\chi_I = \varphi_I^1 - \varphi_I^3$. Note that, when none of the four momenta are parallel to one another, $\chi_I$ is the angle between the two normal vectors created from the two planes spanned by $\{{\bf K}_I^1, {\bf K}_I^2\}$ and $\{{\bf K}_I^3, {\bf K}_I^4\}$. This statement is shown by $|{\bf K}_I^1\times{\bf K}_I^2||{\bf K}_I^3\times{\bf K}_I^4|\cos\left(\chi_I\right) = \left({\bf K}_I^1\times{\bf K}_I^2\right)\cdot\left({\bf K}_I^3\times{\bf K}_I^4\right) = k_I^2K_I^1K_I^3\sin\left(\theta_I^1\right)\sin\left(\theta_I^3\right)\cos\left(\varphi_I^1 - \varphi_I^3\right) = |{\bf K}_I^1\times {\bf k}_I||{\bf K}_I^3 \times {\bf k}_I|\cos\left(\varphi_I^1 - \varphi_I^3\right)$. 
 \subsubsection{Mode-Function Integration}
The time-dependent part is less straightforward. While the form presented in Eq.~\eqref{eq:tvar} is algebraically compact and insightful, each integral diverges by itself as conformal time goes to today (i.e. there is an IR divergence). Only the difference between the two integrals is manifestly convergent. Therefore, combining the two pieces together, 
\begin{align}
{\rm Tvar}_h\left({\bf K}_I\right)&= {\rm Re}\left[\int_{-\infty}^0 \frac{d\tau_1}{(H\tau_1)^2} \mathcal{G}_+\left({\bf K}_I^1, \tau_1\right)\mathcal{G}_+\left({\bf K}_I^2, \tau_1 \right)\mathcal{B}_+^h\left({\bf K}_I^3, {\bf K}_I^4, \tau_1\right)\right]\label{eq:TvarB+}\\
\nonumber \mathcal{B}_+^h\left({\bf K}_I^3, {\bf K}_I^4, \tau_1\right) &= \int_{-\infty}^0 \frac{d\tau_2}{(H\tau_2)^2}\Bigg{[}\mathcal{G}_+({\bf K}_I^3, \tau_2)\mathcal{G}_+\left({\bf K}_I^4, \tau_2\right)u_h\left[{\bf k}_I, {\rm max}(\tau_1, \tau_2)\right]u_h^*\left[{\bf k}_I, {\rm min}(\tau_1, \tau_2)\right] \\
&\hspace{2cm}- \mathcal{G}_-\left({\bf K}_I^3, \tau_2\right)\mathcal{G}_-\left({\bf K}_I^4, \tau_2\right)u_h^*({\bf k}_I, \tau_1)u_h({\bf k}_I, \tau_2)\Bigg{]}.\label{eq:B+}
\end{align}
\subsection{Without dCS}\label{subsec:wo_dCS}
We now reproduce the results of Ref.~\cite{0811.3934} for the GE scalar trispectrum. Evaluating Eq.~\eqref{eq:TvarB+} using Mathematica for inflation without dCS, we obtain 
\begin{align}
\nonumber {\rm Tvar}_\pm\left({\bf K}_I\right)\Big{|}_{\mu = 0} = -\frac{\bar{A}_t}{k_I^3}\frac{H^4}{\left[\prod_{i = 1}^42 k_i^3\right]}\Bigg{\{}&\frac{K_I^1 + K_I^2}{\left[a^I_{34}\right]^2}\left[\frac{1}{2}\left(a^I_{34} + k_I\right)\left(\left[a_{34}^I\right]^2 - 2b_{34}^I\right) + k_I^2(K_I^3 + K_I^4)\right] + \left[(1, 2) \leftrightarrow (3, 4)\right]\\
\nonumber + &\frac{K_I^1 K_I^2}{k_t}\left[\frac{b^I_{34}}{a^I_{34}} - k_I + \frac{k_I}{a^I_{12}}\left(K_I^3 K_I^4 - k_I\frac{b^I_{34}}{a^I_{34}}\right)\left(\frac{1}{k_t} + \frac{1}{a^I_{12}}\right)\right] + [(1, 2)\leftrightarrow (3,4)]\\
- &\frac{k_I}{a^I_{12}a^I_{34}k_t}\left[b^I_{12}b^I_{34} + 2k_I^2\left(\prod_{i = 1}^4 k_i\right)\left(\frac{1}{k_t^2} + \frac{1}{a^I_{12}a^I_{34}} + \frac{k_I}{k_t a^I_{12}a^I_{34}}\right)\right]\Bigg{\}},
\end{align}
with $k_t = \sum_{i = 1}^4 k_i$ the sum of momenta magnitudes, $a_{ij}^I = K_I^i + K_I^j + k_I$, and $b_{ij}^I = (K_I^i + K_I^j)k_I + K_I^i K_I^j$. Therefore, the parity-even GE scalar trispectrum is
\begin{align}
{\rm Re}\langle \delta\phi({\bf k}_1)\delta\phi({\bf k}_2)\delta\phi({\bf k}_3)\delta\phi({\bf k}_4)\rangle_c^{\prime {\rm GE}}\Big{|}_{\mu = 0} &= -\sum_I {\rm Re}\left[\mathcal{P}_+\left({\bf K}_I\right)\right]{\rm Tvar}_+\left({\bf K}_I\right)\Big{|}_{\mu = 0},\\
{\rm Re}\left[\mathcal{P}_+\left({\bf K}_I\right)\right]&= \frac{1}{2}\left[K_I^1 K_I^3 \sin\left(\theta_I^1\right)\sin\left(\theta_I^3\right)\right]^2\cos\left(2\chi_I\right),
\end{align} 
and the parity-odd GE scalar trispectrum is zero. We note no approximations have been made at this point.  

\begin{figure}
\tdplotsetmaincoords{60}{120} 
\begin{tikzpicture} [scale=3, tdplot_main_coords, axis/.style={->,black,thick}, 
vector/.style={-stealth,black,ultra thick}, 
vector guide/.style={dashed, black,thick}]

\coordinate (O) at (0,0,0);


\pgfmathsetmacro{\ax}{1}
\pgfmathsetmacro{\ay}{1}
\pgfmathsetmacro{\az}{1}

\pgfmathsetmacro{\Ox}{0.5\ax}
\pgfmathsetmacro{\Oy}{-0.3\ay}
\pgfmathsetmacro{\Oz}{0.6\az}
\pgfmathsetmacro{\Tx}{-0.4\ax}
\pgfmathsetmacro{\Ty}{1.0\ay}
\pgfmathsetmacro{\Tz}{\az}

\coordinate (k_I) at (0, 0, 0.2\az);
\coordinate (K_I^1) at (\Ox,\Oy, \Oz);
\coordinate (K_I^3) at (\Tx,\Ty,\Tz);

\draw[axis] (0,0,0) -- (1,0,0) node[anchor=north east]{$e_1$};
\draw[axis] (0,0,0) -- (0,1,0) node[anchor=north west]{$e_2$};
\draw[axis] (0,0,0) -- (0,0,1) node[anchor=south]{$e_3$};


\draw[vector, pink!60!gray] (O) -- (k_I);
\draw[vector, blue!50!green] (O) -- (K_I^1);
\draw[vector, red!50!orange] (K_I^1) -- (k_I);
\draw[vector, violet!50!blue]  (k_I) -- (K_I^3);
\draw[vector, orange!50!yellow] (K_I^3) -- (O);

\draw[vector guide]         (K_I^1) -- (\Ox,\Oy,0);
\draw[vector guide]         (O) -- (\Ox,\Oy,0);
\draw[vector guide]         (K_I^3) -- (\Tx,\Ty,0);
\draw[vector guide]         (O) -- (\Tx,\Ty,0);
\draw[vector guide]         (\Ox, 0, 0) -- (\Ox, \Oy, 0);
\draw[vector guide]         (0, \Ty, 0) -- (\Tx, \Ty, 0);

\node[tdplot_main_coords,blue!50!green, anchor=east]
at (0.1\az, -0.2\ay, 0.1\az){${\bf K}_I^1$};
\node[tdplot_main_coords, red!50!orange, anchor=south east]
at (0, -0.1\ay, 0.3\az){${\bf K}_I^2$};
\node[tdplot_main_coords, violet!50!blue, anchor=south]
at (-0.5\ax, 0.2\ay, 0.35\az){${\bf K}_I^3$};
\node[tdplot_main_coords, orange!50!yellow, anchor= north]
at (0, 0.65\ay, 0.6\az){${\bf K}_I^4$};
\node[tdplot_main_coords, pink!60!gray, anchor = west] 
at (0, -0.06\ay, -0.1\az){${\bf k}_I$};
\end{tikzpicture}
\caption{General collapsed trispectrum in channel $I \in \{s, t, u\}$ formed by the four momenta ${\bf K}_I^i, \ i \in \{1, 2, 3, 4\}$ with propagator momentum ${\bf k}_I$. We draw the quadrilateral as non-self-intersecting and scale the cross-products to unit length for ease of visualization.}\label{fig:coll}
\end{figure}
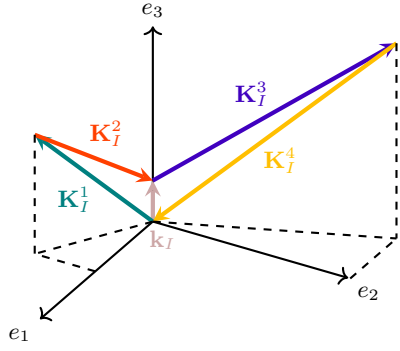

We now turn our attention towards one of three collapsed limits ${\bf k}_i \gg {\bf k}_I \approx 0 $, $i \in \{1, 2, 3, 4\}$, with each limit specified by the channel $I$, see Fig.~\ref{fig:coll}. Each limit also implies the corresponding equalities $K_I^1 \approx K_I^2$ and $K_I^3 \approx K_I^4$. As a result, the time-ordered variance, in the channel of the chosen limit, evaluates to 
\begin{align}
{\rm Tvar}_\pm\left({\bf K}_I\right)\Big{|}^{I-{\rm Coll}.}_{\mu = 0} &= -\frac{\bar{A}_t}{k_I^3}\frac{H^4}{\left[\prod_{i =  1}^4 2k_i^3\right]}\frac{9 K_I^1 K_I^3}{4},
\end{align}
with the other two channels, in the given limit, subdominant. As we will see, the collapsed limit gives significant simplification when we move forward to the parity-odd case. Switching to the curvature perturbation, we then obtain~\footnote{We note that this form is slightly different than Ref.~\cite{0811.3934}, as we retain the factors of $\sin\left(\theta_I^i\right)$. These factors are necessary as the collapsed limit does not specify in which direction ${\bf k}_I$ goes to zero. } 
\begin{align}
{\rm Re}\left\langle\zeta\left({\bf k}_1\right)\zeta\left({\bf k}_2\right)\zeta\left({\bf k}_3\right)\zeta\left({\bf k}_4\right)\right\rangle_c^{\prime {\rm GE}}\Big{|}_{\mu = 0}^{I-{\rm Coll}.} &= \frac{9}{16}\bar{r}\cos\left(2\chi_I\right)\sin^2\left(\theta_I^1\right)\sin^2\left(\theta_I^3\right)P_\zeta\left(k_I\right)P_\zeta\left(K_I^1\right)P_\zeta\left(K_I^3\right).
\end{align}
Since the diagram in Fig.~\ref{fig:4PTGE} has four boundary scalar modes and two bulk graviton modes, the trispectrum is always proportional to two factors of the scalar primordial power spectra and one factor of the primordial tensor power spectrum. The specific angular dependence arises for massless spin-2 particles in de Sitter space and corresponds to associated Legendre polynomials (as opposed to spherical harmonic functions in flat-space)~\cite{2205.01692}. As shown in Ref.~\cite{0811.3934}, the natural smallness, due to the factor of $\bar{r}$, of this trispectrum comes from a consistency condition for the tensor-scalar-scalar bispectrum in the squeezed limit where the tensor mode has a long wavelength. 

\subsection{With Standard dCS}\label{subsec:w_dCS}

In general, due to the presence of the confluent hypergeometric function $U$ in the graviton mode functions, it is difficult to find a closed-form solution for the integral in Eq.~\eqref{eq:TvarB+}. Hence, we use two simplifying assumptions. First, we reassert that we assume a small chemical potential, $\mu \lesssim 1$, for the same reasons as the graviton mode functions and primordial power spectra. Second, we assume that the effect of dCS is only to modify the amplitude of the GE trispectrum, while retaining the same shape, $U(2 \mp i\mu, 4, 2ik\tau)\approx \mathcal{C}_\pm \times U(2, 4, 2ik\tau)\implies \mathcal{C}_\pm = 1/\Gamma\left(2 \mp i\mu\right)$ by matching equality at $k\tau = 0$. 

\begin{figure}
\includegraphics[width = 0.6\textwidth]{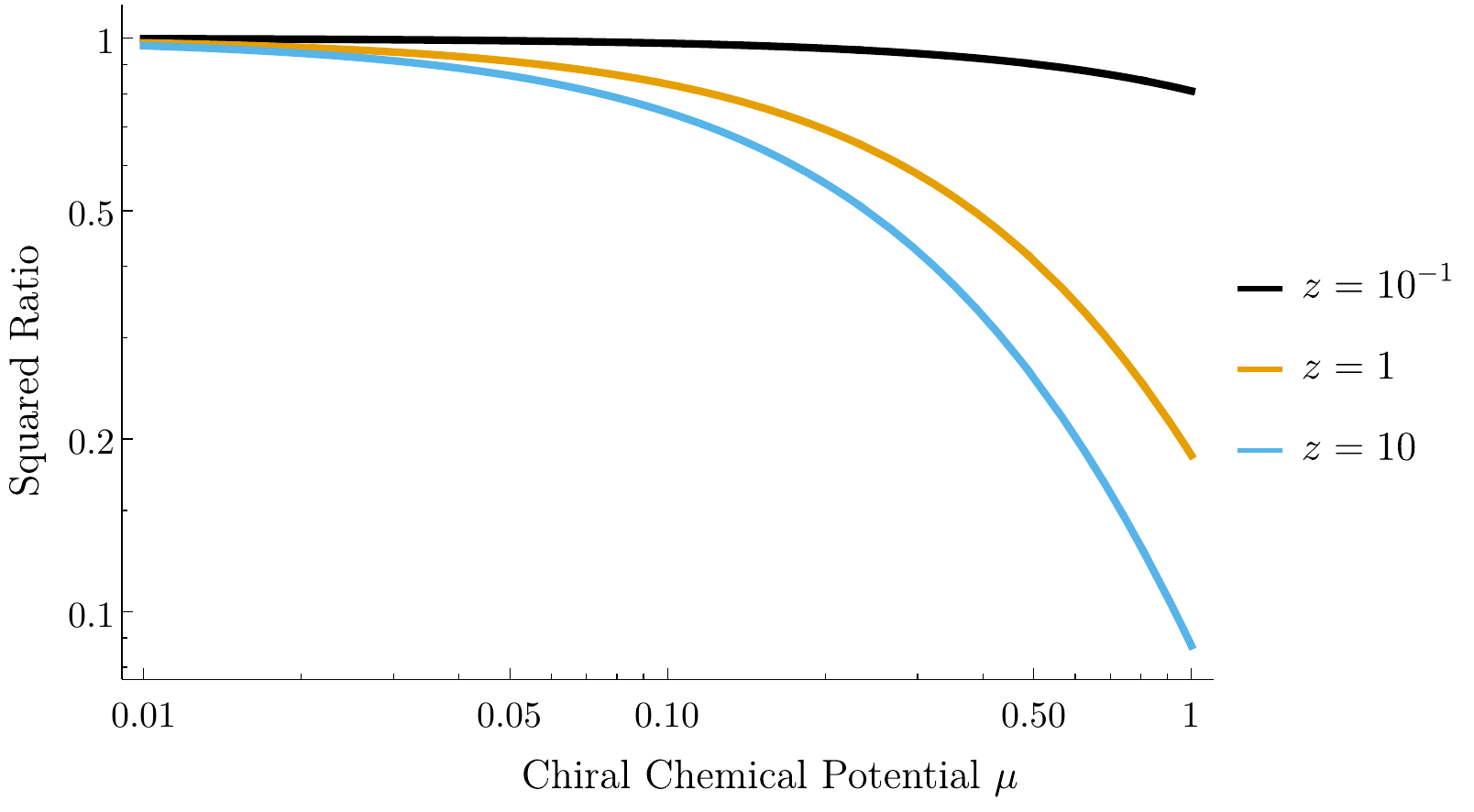}
\caption{The squared ratio between the hypergeometric function approximation $\mathcal{C}_\pm U(2, 4, -2iz)$ and the exact solution $U(2 \mp i \mu, 4, -2iz)$, with $z = -k\tau$ . We take the square as it is what appears in the mode integration of Eq.~\eqref{eq:tvar} up to the different conformal time arguments. For $\mu \lesssim 0.5$, the discrepancy around $z \sim 1$ is about less than order one. }\label{fig:approx_validity}
\end{figure}

In order to see the validity of this approximation, we point out that the dominant contribution of the graviton mode functions to the integrand in Eq.~\eqref{eq:TvarB+} occurs around horizon crossing, $k \tau_i \sim 1, i \in \{1, 2\}$ with $k$ the graviton momentum. Moreover, at sub-horizon values $k\tau_i \gtrsim 1/\mu \gtrsim 1$, when the graviton exhibits a ghost mode, the integrand is highly oscillatory, and hence very subdominant~\cite{hep-th/0506236}. Furthermore, as one goes from $\mu \lesssim 1$ to $\mu \ll 1$, the validity of this approximation increases at much smaller scales, see Fig.~\ref{fig:approx_validity}. As a rule of thumb, this approximation works (i.e. it has $\lesssim \mathcal{O}\left(1\right)$  error) for $\mu \lesssim 0.5$, although for simplicity we will continue to refer to $\mu \lesssim 1$ as our limit. Given the current experimental sensitivity to trispectra in general~\cite{2210.16320}, we consider this approximation reasonable.  Moreover, while the above discussion may seem like our approximation is possibly not optimal for sub-horizon physics, we emphasize that in the collapsed limit, the ghost mode will not contribute and this approximation is in fact exact.
With these two assumptions, we evaluate ${\rm Tvar}$ analytically to obtain
\begin{align}
{\rm Tvar}_\pm\left({\bf K}_I\right)&= \Theta_\pm {\rm Tvar}_\pm\left({\bf K}_I\right)\Big{|}_{\mu = 0},
\end{align}
with corresponding parity-even and -odd trispectra
\begin{align}
\label{eq:main_even}
{\rm Re}\langle \delta\phi({\bf k}_1)\delta\phi({\bf k}_2)\delta\phi({\bf k}_3)\delta\phi({\bf k}_4)\rangle_c^{\prime {\rm GE}} &= \Theta_{\rm even}{\rm Re}\langle \delta\phi({\bf k}_1)\delta\phi({\bf k}_2)\delta\phi({\bf k}_3)\delta\phi({\bf k}_4)\rangle_c^{\prime {\rm GE}}\Big{|}_{\mu  = 0}\\
{\rm Im}\langle \delta\phi({\bf k}_1)\delta\phi({\bf k}_2)\delta\phi({\bf k}_3)\delta\phi({\bf k}_4)\rangle_c^{\prime {\rm GE}} &= -\Theta_{\rm odd}\sum_I {\rm Im}\left[\mathcal{P}_+\left({\bf K}_I\right)\right]{\rm Tvar}_+\left({\bf K}_I\right)\Big{|}_{\mu = 0}\label{eq:main_odd},\\
\nonumber {\rm Im}\left[\mathcal{P}_+\left({\bf K}_I\right)\right] &= \frac{1}{2}\left[K_I^1K_I^3\sin\left(\theta_I^1\right)\sin\left(\theta_I^3\right)\right]^2\sin\left(2\chi_I\right).
\end{align}

Since the dCS dependence has been factored out as an amplitude, we evaluate the collapsed limit in a straightforward fashion,
\begin{align}
\label{eq:Even_Coll}{\rm Re}\left\langle\zeta\left({\bf k}_1\right)\zeta\left({\bf k}_1\right)\zeta\left({\bf k}_3\right)\zeta\left({\bf k}_4\right)\right\rangle_c^{\prime {\rm GE}}\Big{|}_{I-{\rm Coll.}} &= \frac{9}{16}r\cos\left(2\chi_I\right)\sin^2\left(\theta_I^1\right)\sin^2\left(\theta_I^3\right)P_\zeta(k_I)P_\zeta\left(K_I^1\right)P_\zeta\left(K_I^3\right),\\
\label{eq:Odd_Coll}{\rm Im}\left\langle\zeta\left({\bf k}_1\right)\zeta\left({\bf k}_1\right)\zeta\left({\bf k}_3\right)\zeta\left({\bf k}_4\right)\right\rangle_c^{\prime {\rm GE}}\Big{|}_{I-{\rm Coll.}} &= \frac{9}{16}\Pi_{\rm circ}r\sin\left(2\chi_I\right)\sin^2\left(\theta_I^1\right)\sin^2\left(\theta_I^3\right)P_\zeta(k_I)P_\zeta\left(K_I^1\right)P_\zeta\left(K_I^3\right).
\end{align}
Therefore, the ratio of the parity-odd and -even trispectra, in the $I$-collapsed limit, depends only on the degree of gravitational circular polarization and the orientation of the different momenta, 
\begin{align}\label{eq:ratio}
\frac{{\rm Odd}}{{\rm Even}}\Bigg{|}_{I-{\rm Coll.}}= \Pi_{\rm circ}\cot\left(2\chi_I\right). 
\end{align}

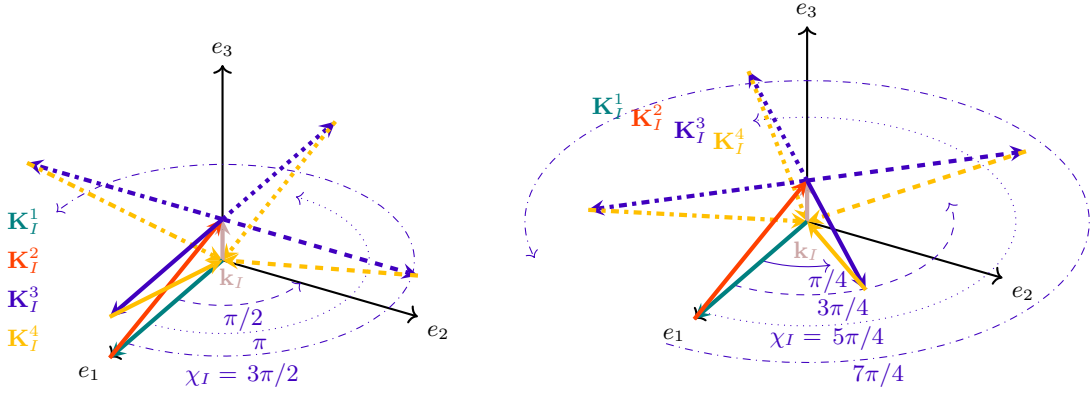
\begin{figure}
\tdplotsetmaincoords{60}{120} 
\begin{tikzpicture} [scale=3, tdplot_main_coords, axis/.style={->,black,thick}, 
vector/.style={-stealth,black,ultra thick}, 
vector guide/.style={dashed, black,thick}]

\coordinate (O) at (0,0,0);


\pgfmathsetmacro{\ax}{1}
\pgfmathsetmacro{\ay}{1}
\pgfmathsetmacro{\az}{1}

\pgfmathsetmacro{\Ox}{\ax}
\pgfmathsetmacro{\Oy}{0}
\pgfmathsetmacro{\Oz}{0}
\pgfmathsetmacro{\TxO}{\az}
\pgfmathsetmacro{\TyO}{0}
\pgfmathsetmacro{\TzO}{0}

\pgfmathsetmacro{\TxTw}{0}
\pgfmathsetmacro{\TyTw}{\ay}
\pgfmathsetmacro{\TzTw}{0}

\pgfmathsetmacro{\TxT}{-\az}
\pgfmathsetmacro{\TyT}{0}
\pgfmathsetmacro{\TzT}{0}

\pgfmathsetmacro{\TxF}{0}
\pgfmathsetmacro{\TyF}{-\ay}
\pgfmathsetmacro{\TzF}{0}

\pgfmathsetmacro{\Iz}{0.2\az}
\coordinate (k_I) at (0, 0, \Iz);
\coordinate (K_I^1) at (\Ox,\Oy, \Oz);
\coordinate (1K_I^3) at (\TxO,\TyO,\TzO + \Iz);
\coordinate (2K_I^3) at (\TxTw,\TyTw,\TzTw + \Iz);
\coordinate (3K_I^3) at (\TxT,\TyT,\TzT + \Iz);
\coordinate (4K_I^3) at (\TxF,\TyF,\TzF + \Iz);

\draw[axis] (0,0,0) -- (1,0,0) node[anchor=north east]{$e_1$};
\draw[axis] (0,0,0) -- (0,1,0) node[anchor=north west]{$e_2$};
\draw[axis] (0,0,0) -- (0,0,1) node[anchor=south]{$e_3$};



\draw[->, dashed, violet!50!blue] (0.4,0,0) arc (0:90:0.4);
\draw[->, dotted, violet!50!blue] (0.65,0,0) arc (0:180:0.65);
\draw[->, dash dot, violet!50!blue] (0.85,0,0) arc (0:270:0.85);

\draw[vector, dashed, violet!50!blue]  (k_I) -- (2K_I^3);
\draw[vector, dashed, orange!50!yellow] (2K_I^3) -- (O);

\draw[vector, dotted, violet!50!blue]  (k_I) -- (3K_I^3);
\draw[vector, dotted, orange!50!yellow] (3K_I^3) -- (O);

\draw[vector, dash dot, violet!50!blue]  (k_I) -- (4K_I^3);
\draw[vector, dash dot, orange!50!yellow] (4K_I^3) -- (O);

\draw[vector, pink!60!gray] (O) -- (k_I);
\draw[vector, blue!50!green] (O) -- (K_I^1);
\draw[vector, red!50!orange] (K_I^1) -- (k_I);

\draw[vector, violet!50!blue]  (k_I) -- (1K_I^3);
\draw[vector, orange!50!yellow] (1K_I^3) -- (O);


\node[tdplot_main_coords,blue!50!green, anchor=east]
at (\ax, -0.3\ay, 0.6\az){${\bf K}_I^1$};
\node[tdplot_main_coords, red!50!orange, anchor=east]
at (\ax, -0.3\ay, 0.4\az){${\bf K}_I^2$};
\node[tdplot_main_coords, violet!50!blue, anchor=east]
at (\ax, -0.3\ay, 0.2\az){${\bf K}_I^3$};
\node[tdplot_main_coords, orange!50!yellow, anchor= east]
at (\ax, -0.3\ay, 0.0\az){${\bf K}_I^4$};
\node[tdplot_main_coords, pink!60!gray, anchor = west] 
at (0, -0.06\ay, -0.1\az){${\bf k}_I$};

\node[tdplot_main_coords, violet!50!blue, anchor=north]
at (0.25\ax, 0.25\ay, 0){$\pi/2$};
\node[tdplot_main_coords, violet!50!blue, anchor=north]
at (0.45\ax, 0.45\ay, 0){$\pi$};
\node[tdplot_main_coords, violet!50!blue, anchor=north]
at (0.6\ax, 0.6\ay, 0){$3\pi/2$};
\node[tdplot_main_coords, violet!50!blue, anchor=north]
at (0.8\ax, 0.4\ay, 0){$\chi_I=$};

\end{tikzpicture}
\qquad
\begin{tikzpicture} [scale=3, tdplot_main_coords, axis/.style={->,black,thick}, 
vector/.style={-stealth,black,ultra thick}, 
vector guide/.style={dashed, black,thick}]

\coordinate (O) at (0,0,0);


\pgfmathsetmacro{\ax}{1}
\pgfmathsetmacro{\ay}{1}
\pgfmathsetmacro{\az}{1}

\pgfmathsetmacro{\Ox}{\ax}
\pgfmathsetmacro{\Oy}{0}
\pgfmathsetmacro{\Oz}{0}

\pgfmathsetmacro{\TxO}{0.7\ax}
\pgfmathsetmacro{\TyO}{0.7\ay}
\pgfmathsetmacro{\TzO}{0}

\pgfmathsetmacro{\TxTw}{-0.7\ax}
\pgfmathsetmacro{\TyTw}{0.7\ay}
\pgfmathsetmacro{\TzTw}{0}

\pgfmathsetmacro{\TxT}{-0.7\ax}
\pgfmathsetmacro{\TyT}{-0.7\ay}
\pgfmathsetmacro{\TzT}{0}

\pgfmathsetmacro{\TxF}{0.7\ax}
\pgfmathsetmacro{\TyF}{-0.7\ay}
\pgfmathsetmacro{\TzF}{0}

\pgfmathsetmacro{\Iz}{0.2\az}
\coordinate (k_I) at (0, 0, \Iz);
\coordinate (K_I^1) at (\Ox,\Oy, \Oz);
\coordinate (1K_I^3) at (\TxO,\TyO,\TzO + \Iz);
\coordinate (2K_I^3) at (\TxTw,\TyTw,\TzTw + \Iz);
\coordinate (3K_I^3) at (\TxT,\TyT,\TzT + \Iz);
\coordinate (4K_I^3) at (\TxF,\TyF,\TzF + \Iz);

\draw[axis] (0,0,0) -- (1,0,0) node[anchor=north east]{$e_1$};
\draw[axis] (0,0,0) -- (0,1,0) node[anchor=north west]{$e_2$};
\draw[axis] (0,0,0) -- (0,0,1) node[anchor=south]{$e_3$};

\draw[->, violet!50!blue] (0.4,0,0) arc (0:45:0.4);
\draw[->, dashed, violet!50!blue] (0.65,0,0) arc (0:135:0.65);
\draw[->, dotted, violet!50!blue] (0.925,0,0) arc (0:225:0.925);
\draw[->, dash dot, violet!50!blue] (1.25,0,0) arc (0:315:1.25);

\draw[vector, dashed, violet!50!blue]  (k_I) -- (2K_I^3);
\draw[vector, dashed, orange!50!yellow] (2K_I^3) -- (O);

\draw[vector, dotted, violet!50!blue]  (k_I) -- (3K_I^3);
\draw[vector, dotted, orange!50!yellow] (3K_I^3) -- (O);

\draw[vector, dash dot, violet!50!blue]  (k_I) -- (4K_I^3);
\draw[vector, dash dot, orange!50!yellow] (4K_I^3) -- (O);

\draw[vector, pink!60!gray] (O) -- (k_I);
\draw[vector, blue!50!green] (O) -- (K_I^1);
\draw[vector, red!50!orange] (K_I^1) -- (k_I);

\draw[vector, violet!50!blue]  (k_I) -- (1K_I^3);
\draw[vector, orange!50!yellow] (1K_I^3) -- (O);


\node[tdplot_main_coords,blue!50!green, anchor=east]
at (\ax, -0.3\ay, \az){${\bf K}_I^1$};
\node[tdplot_main_coords, red!50!orange, anchor=east]
at (\ax, -0.1\ay, \az){${\bf K}_I^2$};
\node[tdplot_main_coords, violet!50!blue, anchor=east]
at (\ax, 0.1\ay, \az){${\bf K}_I^3$};
\node[tdplot_main_coords, orange!50!yellow, anchor= east]
at (\ax, 0.3\ay, \az){${\bf K}_I^4$};
\node[tdplot_main_coords, pink!60!gray, anchor = west] 
at (0.1\ax, -0.05\ay, -0.1\az){${\bf k}_I$};

\node[tdplot_main_coords, violet!50!blue, anchor=north]
at (0.25\ax, 0.25\ay, 0){$\pi/4$};
\node[tdplot_main_coords, violet!50!blue, anchor=north]
at (0.435\ax, 0.435\ay, 0){$3\pi/4$};
\node[tdplot_main_coords, violet!50!blue, anchor=north]
at (0.62\ax, 0.62\ay, 0){$5\pi/4$};
\node[tdplot_main_coords, violet!50!blue, anchor=north]
at (0.86\ax, 0.86\ay, 0){$7\pi/4$};
\node[tdplot_main_coords, violet!50!blue, anchor=north]
at (0.8\ax, 0.4\ay, 0){$\chi_I=$};

\end{tikzpicture}
\caption{Angular configurations that maximize the parity-even (left) and -odd (right) collapsed trispectrum in Eq~\eqref{eq:Even_Coll} and Eq~\eqref{eq:Odd_Coll}, respectively. For simplicity, we hold the orientation of ${\bf K}_I^1$ and ${\bf K}_I^2$ fixed at $\varphi_I^1 = 0$ and allow ${\bf K}_I^3$ and ${\bf K}_I^4$ to rotate together relative to them. The linestyle of the angular arc corresponds to ${\bf K}_I^3$, of the same linestyle, rotated by angle $\varphi_I^3 = \chi_I$. For the parity-even trispectrum, we do not explicitly indicate the rotation by angle $\varphi_I^3 = 0$ when the momenta form a crossed rectangle. Measurements of primordial scalar trispectra that maximize strictly in the above configurations are a direct indication of massless spin-2 exchange, see text for details. We use the same labeling scheme as Fig.~\ref{fig:coll}.
}\label{fig:max_ang}
\end{figure}

The factor of two within the cotangent comes from the expansion of the polarization matrices as a product of two polarization vectors, which, for massless particles, is unique to spin-2 particles~\cite{1607.03735}. Hence, the ratio of the two trispectra encodes direct information about the spin of the graviton, in line with similar findings of Ref.~\cite{2205.01692} for massive particles in oscillatory cosmological collider signals. Massive spin-2 particles with zero chemical potential have five degrees of freedom, while those with a chemical potential have three~\cite{2203.06349}. As a result, their angular dependence will differ~\cite{1811.00024}. We plot the four angular configurations that maximize each collapsed trispectra in Fig.~\ref{fig:max_ang}. Under the assumptions stated at the beginning of the subsection, Eq.~\eqref{eq:main_even} and, in particular, Eq.~\eqref{eq:main_odd} are our main results for the full GE scalar trispectrum in standard dCS inflation, which now contains a parity-odd component. Moreover, Eq.~\ref{eq:Even_Coll} and Eq.~\ref{eq:Odd_Coll} are these results in the collapsed limit, which are exact regardless of our assumptions, and contain the simple relation Eq.~\eqref{eq:ratio} that allows one, in principle, to simply distinguish a parity-odd scalar trispectrum that arises from graviton-exchange from one generated by other sources. 

\subsection{Beyond Standard dCS}\label{subsec:b_dCS}
Due to the aforementioned consistency relation, the trispectra in standard dCS, while non-zero, are unobservably small. However, with modest modifications to the standard setup, one can achieve much larger signals. In this subsection, we quickly demonstrate two such examples: quasi-single field/multi-field dCS inflation and dCS inflation with a superluminal scalar sound speed.  In this vein, we parameterize the trispectrum as 
\begin{align}\label{eq:F}
\langle \zeta({\bf k}_1)\zeta({\bf k}_2)\zeta({\bf k}_3)\zeta({\bf k}_4)\rangle_c \equiv r F S_{\rm GE}^{\rm even}\left({\bf k}_1, {\bf k}_2, {\bf k}_3, {\bf k}_4\right) + i \Pi_{\rm circ}r F S_{\rm GE}^{\rm odd}({\bf k}_1, {\bf k}_2, {\bf k}_3, {\bf k}_4),
\end{align}
with $F$ a trispectrum enhancement factor and  $S_{\rm GE}$ the standard dCS trispectrum from graviton exchange (including momentum-conserving delta functions). In standard inflationary dCS, $F = 1$. In order to make simple connection with the standard calculation, and to prevent any additional complications in the presence of the ghost, we focus solely on the collapsed-limit trispectrum in this subsection. Since in the collapsed configuration the trispectrum can be written exactly as a product of two mixed bispectra, Eq.~\eqref{eq:2bi}, $F^{1/2}$ is the enhancement factor of that bispectrum. We will see that this factor can be larger than $F \gtrsim 10^6$ in the models presented. We note that,  if the enhancement factor is large enough, it is possible to have a small tensor-to-scalar ratio but still a large trispectrum, opening up the possibility of detecting gravitational effects in both LSS and the CMB through their respective trispectra before a direct measurement on $r$ itself. While we discuss only two possibilities in this section,  the factor $F$ can be calculated in a variety of extensions to standard dCS in the collapsed limit and so our parameterization is general. 

\subsubsection{Quasi-Single Field/Multi-Field dCS Inflation}

One potential method for making the graviton-exchange trispectrum to be large is to consider quasi-single field inflation~\cite{0911.3380, 1205.0160}, where the consistency relations for the graviton-exchange scalar trispectra do not need to hold. Refs.~\cite{1407.8204, 1504.05993}, find that the mixed (tensor-scalar-scalar) bispectra obtained lead to a factor $F \sim[(\pi^2/2)w(\nu)]^2 = 10^4 [w(\nu)/20]^2$ larger trispectrum for $\nu = \sqrt{9/4 - (m/H)^2} \lesssim 0.5$, with $m$ the mass of the isocurvature field. However, for $\nu \sim 3/2$, the amplification can be much larger, up to a factor $F \sim N^4$~\cite{0911.3380}, where $N \lesssim 60$ is the number of $e$-folds before the end of inflation. Hence, optimistically, $F \sim 10^7\ (N/60)^4$ for $\nu \sim 3/2$.  We note that this enhancement factor is larger than the $N^2$ effective enhancement factor inferred from Ref.~\cite{0911.3380} because we are considering a tensor-scalar-scalar bispectrum, rather than a scalar-scalar-scalar bispectrum.  

\subsubsection{Superluminal Scalar Sound Speed}
Consider modifying Eq.~\eqref{eq:dCS} by introducing a time-independent scalar sound speed $c_s$,
\begin{align}\label{eq:SL_dCS}
S &= \int d^4x \sqrt{-g}\left[- \frac{1}{2}g^{00}\left(\partial_0\phi\right)\left(\partial_{0}\phi\right) - \frac{1}{2}g^{0i}\left(\partial_0\phi\right)\left(\partial_i\phi\right) - c_s^2 g^{ij}\left(\partial_i\phi\right)\left(\partial_j\phi\right)  + \frac{M_{\rm Pl}^2}{2}R + \frac{\phi}{4f}{^*}RR \right].
\end{align}
Then the scalar primordial power spectrum changes as~\cite{hep-th/9904176, 0709.0293}
\begin{align}
P_\zeta(k) = \frac{1}{c_s}\frac{1}{4\varepsilon k^3}\left(\frac{H}{M_{\rm Pl}}\right)^2 = \frac{2\pi^2 A_s}{k^3}
\end{align}
and the gravitational primordial power spectra remain the same. We now calculate the change in the scalar trispectrum. The additional sound speed introduces a factor of $c_s^4$ from each vertex factor in Fig.~\ref{fig:4PTGE}. We then rescale the conformal time integrals in Eq.~\eqref{eq:tvar} by defining $\tilde{\tau}_i = c_s\tau_i$, after which an additional factor of $c_s^{2}$ comes out. In the collapsed limit, in order to use the correct scalar primordial power spectra, we must add an additional factor of $c_s^3$, however the tensor-to-scalar ratio is then also rescaled by $c_s^{-1}$. As a result, adding all these factors together, we infer than there is an enhancement factor $F = 2\times 10^6 \left(c_s/6\right)^8$ due to the modification in Eq.~\eqref{eq:SL_dCS}. Such superluminal sound speeds can be achieved with non-canonical kinetic terms for the scalar field~\cite{astro-ph/0512066, hep-th/0609150} (see Ref.~\cite{0708.0561} for a discussion regarding the causality of these theories) or in bi-metric gravity~\cite{0803.0859, 0807.1689, 0811.3633, 0908.3898}.

\section{Observables}\label{sec:obs}
Given the dCS-modified primordial scalar trispectra in Sec.~\ref{subsec:w_dCS} and Sec.~\ref{subsec:b_dCS},  we now calculate our present-day observables. In particular, our observables of interest are the parity-odd and -even components of the galaxy trispectrum. With these observables, we then project sensitivities to the amplitude $\Pi_{\rm circ}r$ of the parity-odd trispectrum and the amplitude $r$ of the parity-even trispectrum. Since our calculations for the enhancement factor $F$ were performed in the collapsed limit, we focus here on the sensitivity to such shapes. However, we point out that the estimates in this section are agnostic to how such an enhancement arises.  We refer the reader to Refs.~\cite{astro-ph/9504029, astro-ph/0503375, 1502.00635, 1709.03452, 2107.06287} for a review on how to practically build an optimal estimator to achieve such sensitivities. 

In order to obtain an analytic formula for the sensitivity of galaxy trispectra to these primordial amplitudes, we first consider a simplified setup, somewhat similar to that of Ref.~\cite{2011.09461}, in Sec.~\ref{sec:scale}. We then add complications (i.e. non-linearities and noise) in Sec.~\ref{sec:4cast} in order to perform a forecast for current and upcoming surveys.

\subsection{Sensitivity Scalings}\label{sec:scale}
 We first assume we have a cosmic-variance-limited experiment that measures only linear Fourier modes between $k_{\rm min}$ and $k_{\rm max}$ in a comoving volume $V$. Here, $V$ is a quantity independent from either wavenumber, except for the requirement that the minimum wavenumber does not correspond to a scale larger than the probed volume, $V \gtrsim \left(2\pi/k_{\rm min}\right)^3$. Since we are in the linear regime, the primordial trispectra and power spectra are related to their present-day galaxy values by the same constant factor. Hence, when doing signal-to-noise estimates on galaxy trispectra amplitudes, we can perform the estimation using only primordial quantities. Therefore, the squared signal-to-noise for $r\times F$ from the parity-even sector is~\cite{1502.00635},
\begin{align}\label{eq:PE_SNR}
\left(\frac{S}{N}\right)_{r F}^2 &= \frac{\left(r F\right)^2}{24}\int_{k_{\rm min}}^{k_{\rm max}}\left[\prod_{i = 1}^4\frac{d^3{\bf k}_i}{(2\pi)^3}\right]\frac{[S_{\rm GE}^{\rm even}({\bf k}_1, {\bf k}_2, {\bf k}_3, {\bf k}_4)]^2}{P(k_1)P(k_2)P(k_3)P(k_4)}\\
&= \frac{\left(r F\right)^2}{24}N_{\rm modes}\mathcal{N}_{\rm GE}, 
\end{align}
where $N_{\rm modes} = k_{\rm max}^3 V/(6\pi^2)$ is the number of modes within the volume $V$ and $\mathcal{N}_{\rm GE}$ is the wavenumber integral for graviton exchange. In the collapsed limit, this integral is  $\mathcal{N}_{\rm GE} \approx \sum_I \mathcal{N}_{\rm GE}^{I-{\rm Coll.}}$, with
\begin{align}\label{eqref:N_coll}
\nonumber \mathcal{N}_{\rm GE}^{I-{\rm Coll}.} &= \frac{81}{256}\frac{1}{N_{\rm modes}}\int \frac{d\Omega_{{\bf k}_I}}{(2\pi)^3} k_{\rm min}^3 P^2_\zeta(k_{\rm min})\int_{k_{\rm min}}^{k_{\rm max}}\left[\frac{d^3{\bf K}_I^1 d^3{\bf K}_I^3}{(2\pi)^6}\right]\sin^4\left(\theta_I^1\right)\sin^4\left(\theta_I^3\right)\cos^2\left[2(\varphi_I^1 - \varphi_I^3)\right]\\
&= \frac{3}{200}\frac{k^3_{\rm max}}{k^3_{\rm min}}A_s^2, 
\end{align}
where we relabeled the four momenta integrals by ${\bf K}_I^i$, used the delta function on ${\bf K}_I^4$ and the identity $(2\pi)^3\delta_D^{(3)}(0) = V$ (which appears in $N_{\rm modes}$), swapped integration of ${\bf K}_I^2$ for ${\bf k}_I$, and identified $\int_{I-{\rm Coll}.} dk_I k_I^2 P^2_\zeta(k_I)\approx k_{\rm min}^3 P^2_\zeta(k_{\rm min})$. The collapsed limit enhances the squared SNR from $N_{\rm modes}$ to $N_{\rm modes}(k_{\rm max}/k_{\rm min})^3$ and is identical for all three channels. The cubic dependence on the ratio of maximum and minimum wavenumber is characteristic of a tree-level collapsed trispectrum with a massless mediator, see Eq.~(3.19) of Ref.~\cite{2011.09461}. The estimation for the parity-odd amplitude $\Pi_{\rm circ}r F$ is exactly the same, with the swap of $S_{\rm GE}^{\rm even} \rightarrow S_{\rm GE}^{\rm odd}$ and the corresponding amplitudes in Eq.~\eqref{eq:PE_SNR}, as the angular integrals over $\varphi_I^i$ are equal. Therefore, it follows that, for a dCS signal greater than about $n\sigma$,
\begin{align}\label{eq:amp_SNR}
\left[1 + \delta_{\rm odd}^{P}\left(\Pi_{\rm circ} - 1\right)\right]r \gtrsim 0.04 \left(\frac{n}{3}\right)\left(\frac{8\times 10^5}{F}\right)\left(\frac{10^6}{N_{\rm modes}}\right)^{1/2}\left(\frac{k_{\rm min}}{0.003\ h/{\rm Mpc}}\right)^{3/2}\left(\frac{0.3\ h/{\rm Mpc}}{k_{\rm max}}\right)^{3/2},
\end{align} 
with $P \in \{{\rm even}, {\rm odd}\}$ and we point out once more this estimate holds only for mildly or fully collapsed shapes. We use $0.04$ as our benchmark amplitude for a detection at $3\sigma$ as the degree of gravitational circular polarization must obey $|\Pi_{\rm circ}| \leq 1$ by definition  and the tensor-to-scalar ratio must be $r \lesssim 0.04$ to not exceed existing constraints on tensor modes~\cite{2203.16556}.  For the parity-odd amplitude, we translate Eq.~\eqref{eq:amp_SNR} into a restriction on the dCS decay constant using Eq.~\eqref{eq:g_circ},
\begin{align}\label{eq:f_SNR}
 f \lesssim 4\times 10^{9}\ {\rm GeV}\left(\frac{3}{n}\right)\left(\frac{F}{8\times 10^5}\right)\left(\frac{\varepsilon}{10^{-2}}\right)^{3/2}\left(\frac{H}{10^{14}\ {\rm GeV}}\right)^2\left(\frac{N_{\rm modes}}{10^6}\right)^{1/2}\left(\frac{0.003\ h/{\rm Mpc}}{k_{\rm min}}\right)^{3/2}\left(\frac{k_{\rm max}}{0.3\ h/{\rm Mpc}}\right)^{3/2},
\end{align}
noting that, if the primordial scalar amplitude obeys the relation in Eq.~\eqref{eq:PS}, one can trade the Hubble scale of inflation for the slow-roll parameter, or vice versa, via $H = 10^{14}\ {\rm GeV}\left(\varepsilon/10^{-2}\right)^{1/2}$. We also point out that one should take care to restrict to $\mu \lesssim 1$ when using the above equation so as not to excite the ghost. 

\subsection{Forecasts}\label{sec:4cast}
The presence of both non-linearities and noise in  actual clustering data will cause a degradation of the signal in Eq.~\eqref{eq:amp_SNR}. Therefore, the quantification of both effects are needed to make contact with real experiments, of which we consider both spectroscopic galaxy surveys and 21-cm experiments. In order to capture these effects, while still maintaining the simple analytical form of Eq.~\eqref{eq:amp_SNR}, we make the phenomenological replacement of the cosmic-variance linear approximation $N_{\rm modes} = k_{\rm max}^3 V/(6\pi^2)$ with the more accurate number of modes sensitive to the linear galaxy power-spectrum amplitude~\citep{2106.09713}
\begin{align}\label{eq:P_modes}
N_{\rm modes} &= V\int_{k_{\rm min}}^{k_{\rm max}} \frac{d^3 {\bf k}}{(2\pi)^3}{\rm W}\left({\bf k}\right)\frac{\left[G^2({\bf k}, \bar{z})P_{\rm L}^{gg}({\bf k}, \bar{z})\right]^2}{\left[P_{\rm NL}^{gg}\left({\bf k}, \bar{z}\right) + \bar{n}^{-1}\right]^2}.
\end{align}
 In this replacement, we are considering an experiment that probes redshifts between $z_{\rm min}$ and $z_{\rm max}$ and has average redshift $\bar{z}$. The probed comoving volume is then $V = (4/3)\pi f_{\rm sky}[\chi^3(z_{\rm max}) - \chi^3(z_{\rm min})]$, with $\chi = \int dt/a$  the comoving distance. Within this volume, there exists both a linear, $\delta^g_{\rm L}\left({\bf k}, z\right)$, and non-linear, $\delta^g_{\rm NL}\left({\bf k}, z\right)$, galaxy density field with corresponding linear, $P_{\rm L}^{gg}({\bf k}, z)$, and non-linear, $P_{\rm NL}^{gg}({\bf k}, z)$, galaxy power spectra. For the purpose of forecasting, we assume the Kaiser form for the linear and non-linear power spectra~\cite{Kaiser:1987qv}, 
\begin{align}
P_{a}^{gg}\left({\bf k}, z\right) = \left[b(z) + f(z) \mu^2\right]^2D^2(z)P_{a}^m\left(k\right),
\end{align}
with $a \in \{{\rm L}, {\rm NL}\}$, $b(z)$ the sample bias at redshift $z$, $\mu = \hat{\bf k}\cdot \hat{\bf n}$ the cosine of the angle between the wavenumber and the line of sight  $\hat{\bf n}$, $D(z)$ the linear growth factor (normalized to unity today), $f(z) = d\log D(z)/d\log a(z)$ the linear growth rate, and $P^m_a(k)$ the matter power spectrum today. The linear and non-linear galaxy perturbations are also not fully correlated, which is quantified by the decorrelation propagator $G\left({\bf k}, z\right) \equiv \left\langle \delta^g_{\rm NL}\left({\bf k}, z\right)\delta^g_{\rm L}\left({\bf k}, z\right)\right\rangle_c/\left\langle \delta^g_{\rm L}\left({\bf k}, z\right)\delta^g_{\rm L}\left({\bf k}, z\right)\right\rangle_c$. This propagator is typically modelled by the Gaussian
\begin{align}
G\left({\bf k}, z\right) &\simeq \exp\left[-\frac{1}{2}\left(k_\perp^2 + k_{||}^2\left[1 + f(z)\right]^2\right)\Sigma^2(z)\right], 
\end{align}
with $k_\perp = k\sqrt{1 - \mu^2}$ $\ \left(k_{||} = k\mu\right)$ the wavenumber perpendicular (parallel) to the line of sight, 
\begin{align}
\Sigma(z) = \int_0^\infty \frac{dk}{6\pi^2}P_{\rm L}^m(k, z)
\end{align}
the root-mean-square (RMS) displacement in the Zel'dovich approximation~\cite{1401.5466}, and $P_{\rm L}^m(k, z) = D^2(z)P_{\rm L}^m(k)$ the linear matter power spectrum at redshift $z$. The RMS displacement also characterizes the scale at which non-linearities in the matter (and thus galaxy) field begin to develop, $k_{\rm NL}(z) \sim \Sigma^{-1/2}(z)$, and so we identify this scale as the maximum wavenumber through its evaluation at the average redshift of the sample, $k_{\rm max} = \Sigma^{-1/2}(\bar{z})$. We let the minimum wavenumber $k_{\rm min}$ be determined by the experiment in question. However, for $21$-cm experiments there exists large foregrounds on large scales along the line of sight, and so a wedge ${\rm W}\left({\bf k}\right)$ is required~\cite{1709.06752} to remove these modes from the analysis as they cannot be recovered,
\begin{align}
{\rm W}\left({\bf k}\right) &= \Theta \left\{k_{||} - 
 {\rm max}\left[k_\perp\frac{\chi(z)H(z)}{(1 + z)}\sin\left[\theta_w(z)\right], k_{|| {\rm min}}\right]\right\},\quad \theta_w(z) = N_w\frac{1.22}{2\sqrt{0.7}}\frac{\lambda^{21}_{\rm obs}(z)}{D_{\rm phys}},
\end{align}
with $\Theta(x)$ the Heaviside function, $\theta_w$ the beam width, $\lambda_{\rm obs}^{21} = 21\ {\rm cm}(1 + z)$ the wavelength of an observed $21$-cm line at redshift $z$, and $D_{\rm phys} = 6\ {\rm m}$  the physical diameter of the radio dish for both PUMA and HIRAX. Optimistically, all modes outside the beam width can be recovered, so that $N_w = 1$, but pessimistically a cut of $N_w = 3$ is likely required~\cite{1810.09572}. In the optimistic (pessimistic) case, we also set the minimum wavenumber along the line of sight to be $k_{|| {\rm min}} = 0.01\ (0.1)\ h/{\rm Mpc}$. For galaxy surveys, we set ${\rm W}({\bf k}) = 1$. 

In addition to the non-linearities, there is also intrinsic noise, specified by the source density $\bar{n}(z)$ or its spatial average $\bar{n}$. For spectroscopic galaxy surveys, this noise is the average number density of the sample, while for 21-cm experiments, it is the sum of both thermal and shot noise. For compact notation, we denote the quantity $b(\bar{z})$ to be the average bias weighed by the number density, $\bar{b}(z) = \int dz\,\bar{n}(z) b(z)/\int dz\,\bar{n}(z)$. 
We note that, in the limit of zero non-linearities, infinite $\bar{n}$, and no foregrounds, Eq.~\eqref{eq:P_modes} reduces to $N_{\rm modes} = k_{\rm max}^3 V/(6\pi^2)$. 

Using this mode number replacement, we now roughly calculate the number of modes measured by a variety of upcoming and future experiments in Table~\ref{tab:exp} and estimate how large the amplification factors $F$ must be in order to see a detection in those experiments using Eq.~\eqref{eq:amp_SNR}. We use the biases and number densities listed in Ref.~\cite{2106.09713}, which also contains a more accurate and thorough treatment of different experimental designs, and use \texttt{CLASS} with \texttt{halofit} to calculate linear and non-linear matter power spectra today. We find that current, upcoming, and future experiments are only sensitive to collapsed dCS models with at least an enhancement of around $F \sim 10^{5}-10^6$  at $3\sigma$.
\begin{table}
\centering
\begin{tabular}{|c||c|c|c|c|c||c|c|c|}
 \hline
 \multicolumn{9}{|c|}{Experiments} \\
 \hline
 Name & $[z_{\rm min}, z_{\rm max}]$ &  $f_{\rm sky}$ & $[k_{\rm min}, k_{\rm max}]$ $[h/{\rm Mpc}]$ & $\bar{n}\ [(h/{\rm Mpc})^3] $ &  $\bar{b}$ & $V$ [$({\rm Gpc}/h)^3$] & $N_{\rm modes}$ & $F$\\
 \hline
 BOSS\footnote[1]{Spectroscopic Galaxy Survey}~\cite{1909.05277, 2112.04515} & $[0.43, 0.7]$    & $0.23$  & $[0.01,0.24]$ & $3\times 10^{-4}$ & 2.0  & 3.6 & $9\times 10^4$ & $2\times 10^7$\\
DESI$^a$~\cite{1611.00036} &   $[0.6, 1.7]$  & $0.34$ & $[0.003, 0.31]$  & $3.8\times 10^{-4}$  & 1.1 & 45 & $4\times 10^5$ & $10^6$\\
 Euclid$^a$~\cite{1606.00180} & $[0.9, 1.8]$    & $0.36$  & $[0.003, 0.34]$  & $4.3\times 10^{-4}$ & 1.5 & 44 & $9\times 10^5$ & $7\times 10^5$\\
 MegaMapper$^a$~\cite{1907.11171, 2209.04322} & $[2, 5]$ & $0.34$ & $[0.003, 0.64]$  & $2.5\times 10^{-4}$ & 3.9 & 155 & $6\times 10^6$ & $10^5$\\
 MSE$^a$~\cite{1904.04907} &  $[1.6, 4]$  & $0.24$ & $[0.003, 0.54]$  & $2.9\times 10^{-4}$ & 3.7 & 91 & $5\times 10^6$ & $2\times 10^5$\\
SPHEREx$^a$~\cite{1412.4872, 1606.07039} & $[0.1, 3.0]$ & $0.65$ & $[0.003, 0.37]$  & $3.1\times 10^{-3}$ & 0.65 & 230 & $4\times 10^6$ & $3\times 10^5$\\
\hline
 HIRAX\footnote[2]{21-cm Experiment}~\cite{1607.02059}&  $[0.8, 2.5]$  & $0.36$  & $[(0.01, 0.1), 0.38]$  & $10^{-3}$ & $1.8$  & 88 & $2\times 10^6$ & $\left(2\times 10^6, 8\times 10^7\right)$\\
 PUMA-32K$^b$~\cite{1907.12559}  &  $[2, 6]$  & $0.5$ & $[(0.01, 0.1), 0.71]$  & $7.6\times 10^{-3}$ & $3.2$ & 290 & $4\times 10^7$ & $\left(2\times 10^5, 7\times 10^6\right)$\\
 \hline
\end{tabular}
\caption{List of experimental parameters in order to roughly estimate the number of linear modes $N_{\rm modes}$ and sensitivity to the trispectrum enhancement factor $F$ using Eq~\eqref{eq:P_modes} with Eq.~\eqref{eq:amp_SNR}. $z_{\rm min}$ $(z_{\rm max})$ is the minimum (maximum) observed redshift, $f_{\rm sky}$ is the observed fraction of the sky, and $k_{\rm min}$ ($k_{\rm max}$) is the minimum (maximum) observed wavenumber. We set the maximum wavenumber to the non-linear scale evaluated at the average sample of the redshift, $k_{\rm max} = k_{\rm NL}(\bar{z})$. $\bar{b}$ is the number-density-averaged bias, and $\bar{n}$ the average noise density over the entire redshift range.  The total observed comoving volume is $V = (4/3)\pi f_{\rm sky}[\chi^3(z_{\rm max}) - \chi^3(z_{\rm min})]$, with $\chi = \int dt/a$  the comoving distance calculated using $\Lambda$CDM parameters~\cite{1807.06209} (e.g. $h = 0.674$). The last column contains values of the enhancement factor $F$ needed to detect an $r$ $(r \Pi_{\rm circ})$ value of $0.04$ at $3\sigma$ in the parity-even (-odd) galaxy trispectrum for the experiment in the same row. As a result, all upcoming experiments are only sensitive to collapsed models with at least an enhancement of around $F \sim 10^{5}-10^6$ at $3\sigma$. For 21-cm experiments, we consider two values of $k_{\rm min}$, corresponding to pessimistic and optimistic foreground removal. In these cases, we also change the values of both $N_w$ and $k_{|| {\rm min}}$ accordingly, see text for details. While 21-cm experiments have a comparable, or greater, number of modes, the presence of foregrounds limits the smallest available wavenumber, and therefore decreases the cubic  $(k_{\rm max}/k_{\rm min})^{3/2}$ enhancement. We also note that this cubic scaling is not strictly followed due to the presence of the wedge.}\label{tab:exp}
\end{table}

\section{Discussion}\label{sec:disc}
We present a technical discussion on a variety of assumptions throughout the paper and give a few comments. For a more general overview, we point the reader to the conclusion in Sec.~\ref{sec:conc}. Specifically, we clarify five assumptions and provide four comments. First, we assumed that inflation occurs through the breaking of time-translations. The breaking of spatial translations, through solid inflation~\cite{astro-ph/0404548, 1210.0569, 1306.4160}, will lead to qualitatively and quantitatively  parity-breaking different signatures, e.g. through the presence of anisotropic shapes.

Second, we assumed that the pseudoscalar coupled to the Pontryagin density in Eq.~\eqref{eq:dCS} was the inflaton. This choice can be lifted, and instead the coupling can be with a separate field during inflation.

Third,  we assumed the Bunch-Davies vacuum as an initial condition. This choice too can be lifted, and non-vacuum initial conditions can be considered~\cite{0710.1302, 1110.4688, 1212.1172, astro-ph/9904167}. These initial conditions can lead to large enhancements while also taking some of their maximum values in distinct flattened shapes~\cite{hep-th/0605045}.

Fourth, we required $\mu \lesssim 1$  to prevent both the production of ghost-like graviton modes and the overproduction of tensor modes. Ghost-like modes in the context of scalar inflation have indeed been thoroughly investigated~\cite{hep-th/0312100, astro-ph/0405356, 2210.16320}. In addition, the tensor-to-scalar ratio can be made small by lowering the Hubble scale of inflation. As a result, it is potentially viable to consider, from an EFT perspective, ghost gravitational Chern-Simons inflation. However, it is worth noting that such models require a large hierarchy between the EFT cutoff and the dCS decay constant~\cite{1208.4871}. We leave the investigation of ghost modes in Chern-Simons theories for future work. 

Fifth, we assumed the effect of dCS is to only change the amplitude of the trispectrum and not its shape. While this approximation becomes exact in the collapsed limit, changes in shape are in fact present at smaller scales, as the confluent hypergeometric is not a constant over all ${\bf k}$. In order to evaluate such dependence, the full~\cite{1906.12302} or partial~\cite{2208.13790} Mellin-Barnes formulation could potentially be used. We leave investigations to the precise shape and detectability to such signals for future work. 

A key point in a detectable signal from graviton-exchange is the enhancement factor of the trispectrum. In order to obtain large enhancements, it must be that de Sitter boosts are broken in some manner~\cite{2004.09587}. While we have explicitly considered two possibilities for generating such an enhancement (namely additional scalar field content and superluminal scalar sound speed) it is not true that this list is exhaustive. For example, one could entertain small graviton sound speeds. Ref.~\cite{2109.10189} has delineated a host of methods of obtaining large non-Gaussianities for graviton correlators, and it would be interesting to see how they map on quantitatively onto various enhancement factors. 

We caution that our forecasts for LSS are not exact, in particular for the parity-even case (e.g. we did not take into account non-linear growth, redshift evolution, or the Alcock-Paczynski effect).
We also point out the CMB  does not suffer from these limitations so searches for parity-violation there would be complementary.

Given that we have calculated the parity-violating trispectrum for a massless spin-2 particle, it is obvious to ask whether the same occurs for massless spin-1 exchange as well. To be concrete, consider a $U(1)$ theory with gauge field $A_\mu$ and Chern-Simons coupling to the inflaton.  The main differences are that a) massless spin-1 particles have zero conformal mass and thus their mode functions without Chern-Simons go as $e^{-ik\tau}/\sqrt{k}$ b) the $U(1)$ theory exhibits no ghosts~\footnote{We thank Eiichiro Komatsu for pointing this feature out in conversation.} due to the smaller number of spatial derivatives in the Chern-Simons density and c) the vertex interaction is no longer given by a standard kinetic term. Possibly the simplest vertex interaction term would be an operator of the form $\dot{\phi}A^i \partial_i \phi$. With this setup, the same calculation that we have done can be carried out to calculate its contribution to the scalar parity-odd trispectrum. 

Aside from the scalar trispectrum, dCS is also of interest because it can lead to leptogenesis~\cite{hep-th/0410230, 1711.04800} and thus be a source of the baryon asymmetry today. The necessary ingredients for such a process to occur in our setup are exactly the same, minus the kinetic term vertex interaction in Eq.~\eqref{eq:kin}. Therefore, in principle, one can probe leptogenesis-generating mechanisms using LSS and CMB measurements. One caveat is that the original setup in Ref.~\cite{hep-th/0410230} requires $f \sim M_{\rm Pl}$. However, broadband parametric resonance during preheating can cause an exponential amplification of number densities generated during inflation~\cite{1405.4288, 1508.00891}, thus potentially allowing for smaller values of $f$ to generate smaller asymmetries that are then later amplified. Downstream effects of leptogenesis can also generate interesting non-Gaussian signals, as shown by Ref.~\cite{2112.10793}. Moreover, given the setup of $U(1)$ Chern-Simons inflation one can also obtain baryogenesis~\cite{1107.0318, 1905.13318, 1508.00881}. We leave a detailed investigation into the connections between baryogenesis and both LSS and CMB signals for future work.    

Other extensions to our calculation include calculating the associated real-space galaxy four-point function and both the CMB temperature and polarization trispectra. Since the trispectrum and its real-space counterpart contain the same information, we expect our sensitivity analysis to hold. However, how this signal imprints on the CMB can be very different and so a separate analysis is required, which we leave for future work. We note that late-time parity-violation signals from Chern-Simons gravity are generically suppressed on large scales~\cite{2210.16320}.
\section{Conclusion}\label{sec:conc}
In this paper, we have calculated the primordial scalar trispectrum of dynamical Chern-Simons gravity (dCS) in three variants (one standard, two beyond) and estimated the sensitivity of LSS surveys to their amplitudes. In order to do so, we first reviewed the phenomenology of standard inflationary dCS in Sec.~\ref{sec:th}, showing how one graviton polarization is amplified due to a tachyonic instability that depends on the chiral chemical potential $\mu \lesssim 1$, whose upper bound is to prevent the excitation of a ghost mode. In other words, this chemical potential yielded a nonzero degree of gravitational circular polarization $\Pi_{\rm circ} \propto \mu$. 

Then, in Sec.~\ref{sec:in-in}, we calculated the trispectrum due to graviton-exchange between two pairs of scalars using Schwinger-Keldysh (SK) diagrammatic rules, presented in Sec.~\ref{subsec:gen_calc}. This trispectrum exists in inflation without dCS, and so we first reviewed this case in Sec.~\ref{subsec:wo_dCS}. Then we turned on the standard dCS coupling in Sec.~\ref{subsec:w_dCS}, finding a non-zero parity-odd scalar trispectrum, Eq.~\eqref{eq:main_odd} and Eq.~\eqref{eq:Odd_Coll}, whose dependency on dCS entered through the degree of gravitational circular polarization, Eq.~\eqref{eq:g_circ}. Within dCS, this parity-odd trispectrum is the dominant parity-odd signal of inflation and our calculation is the first example of parity-odd scalar trispectra arising from massless spin exchange.  We also found corrections to the standard parity-even scalar trispectra in Eq.~\eqref{eq:main_even} and Eq.~\eqref{eq:Even_Coll}  which are $\propto \mu^2$. Interestingly, we also found that the ratio of the odd trispectrum to the even trispectrum took a simple form in the collapsed limit, Eq.~\eqref{eq:ratio}. That is, this ratio depended only on the degree of gravitational circular polarization, the graviton's spin, and the orientation of trispectra momenta. As a result, it is then possible, in principle, to simply distinguish a parity-odd trispectrum that arises from graviton exchange from one that is generated by other mechanisms. 

In the standard case, both the parity-even and parity-odd scalar trispectra are small due a consistency condition for the associated parity-even tensor-scalar-scalar bispectrum when the tensor mode has a long wavelength, as shown in Ref.~\cite{0811.3934}. However, with a modest change to the standard dCS setup, we showed, in Sec.~\ref{subsec:b_dCS}, that it is possible to generate much larger trispectra. Specifically, we considered two variants beyond the standard setup: one where the dCS term enters into quasi-single field/multi-field inflation and one where dCS inflation occurs with a superluminal scalar sound speed. In both cases, we showed that, in the collapsed limit, it is possible to get scalar trispectrum enhancements by a factor $F \gtrsim 10^6$. Our parameterization of this enhancement  allows for straightforward extension to other models beyond standard dCS. If the enhancement factor is large enough, it is possible to detect gravitational effects in scalar trispectra before a direct measurement on $r$ itself. 

Given the ability to make large trispectra, we translated our primordial calculations to the observed galaxy trispectrum in Sec.~\ref{sec:obs}.  We first found that a hypothetical cosmic-variance-limited LSS survey measuring $N_{\rm modes}$ linear modes between $k_{\rm min}$ and $k_{\rm max}$  has sensitivity to parity-even and -odd amplitude values according to Eq.~\eqref{eq:amp_SNR}. Translating this estimate into a statement on dCS, we found that such a survey would be sensitive to very large dCS decay constants, $f \gtrsim 10^9\ {\rm GeV}$ [see Eq.~\eqref{eq:f_SNR} for the general formula] that have not been able to be probed by experiments thus far~\cite{0708.0001, 1705.07924,  2101.11153, 2103.09913, 2208.02805}. In order to account for the effects of nonlinearities and noise, we made a phenomenological replacement of $N_{\rm modes}$ by way of Eq.~\eqref{eq:P_modes}, allowing us to use our previous sensitivity formula. With this replacement, we connected our sensitivity estimates to current and upcoming experiments in Table~\ref{tab:exp}. We found that current and future LSS surveys are sensitive to collapsed dCS models with amplification factors of at least $F\sim 10^5 - 10^6$ at $3\sigma$, which are well within the two examples we demonstrated in Sec.~\ref{subsec:b_dCS}. 

We discussed several technical assumptions and extensions in Sec.~\ref{sec:disc}, pointing out that a similar signal is present within the CMB and that the very large dCS decay constants probed by LSS (and hence the CMB) can potentially yield leptogenesis. We therefore conclude by stating that the parity-odd scalar trispectrum is a valuable tool to study inflation, gravitational parity violation, as well as potentially baryogenesis. 

\subsection*{Acknowledgments}
C.C.S would like to thank the Aspen Center for
Physics, which is supported by National Science Foundation grant PHY-1607611, for hospitality and inspiring this project. In addition, C.C.S would like to thank David Stefanyszyn, Ken Van Tilburg, and  Jiamin Hou  for comments on a early draft, as well as Giovanni Cabass and Eiichiro Komatsu for useful discussions. O.H.E.P. is a Junior Fellow of the Simons Society of Fellows and thanks the Simons Foundation for support and blackboard quality.  MK was supported by NSF Grant No.\ 2112699 and the Simons Foundation. SA was supported by the Simons Foundation award number 896696.

\end{document}